\documentclass[11pt]{article}
\usepackage[final]{acl}
\usepackage[utf8]{inputenc}
\usepackage{times}

\usepackage[T1]{fontenc}
\usepackage[utf8]{inputenc}

\usepackage{microtype}
\usepackage{latexsym}
\usepackage{amsmath,amssymb,amsthm}

\usepackage{graphicx}
\usepackage{caption}
\usepackage{subcaption}
\usepackage[font=small,labelfont=bf]{caption} 

\usepackage{booktabs,multirow,array}
\usepackage{colortbl}
\usepackage{xcolor}

\usepackage{enumitem}
\usepackage{url}
\usepackage{hyperref}
\usepackage{float}
\usepackage{stfloats}

\usepackage{inconsolata}
\usepackage{verbatim}

\usepackage{etoolbox}
\usepackage{adjustbox}
\usepackage{pgf}

\usepackage{algpseudocode}

\usepackage{tikz}
\usetikzlibrary{shapes,arrows.meta,positioning}

\newlength{\subfigsep}
\newcommand\blfootnote[1]{%
  \begingroup
  \renewcommand\thefootnote{}%
  \footnotetext{#1}%
  \addtocounter{footnote}{-1}%
  \endgroup
}

\title{\textbf{\texttt{Prosody as Supervision:}} Bridging the Non-Verbal–Verbal for Multilingual Speech Emotion Recognition}

\author{
\textbf{Girish\textsuperscript{1}\thanks{Equal contribution as a first author.}} \quad
\textbf{Mohd Mujtaba Akhtar\textsuperscript{2}\footnotemark[1]} \quad
\textbf{Muskaan Singh\textsuperscript{3}\thanks{Corresponding author.}}\\
\textsuperscript{1}UPES, India \quad
\textsuperscript{2}Veer Bahadur Singh Purvanchal University, India \quad
\textsuperscript{3}Ulster University, UK\\
\texttt{\{girish.research.pr,mmakhtar.research\}@gmail.com}, \quad \texttt{m.singh@ulster.ac.uk}
}

\begin{document}
\maketitle
\begin{abstract}
In this work, we introduce a paralinguistic-supervision paradigm for low-resource multilingual speech emotion recognition (LRM-SER) that leverages non-verbal vocalizations to exploit prosody-centric emotion cues. Unlike conventional SER systems that rely heavily on labeled verbal speech and suffer from poor cross-lingual transfer, our approach reformulates LRM-SER as non-verbal-to-verbal transfer, where supervision from a labeled non-verbal source domain is adapted to unlabeled verbal speech across multiple target languages. To this end, we propose \textbf{NOVA-ARC}, a geometry-aware framework that models affective structure in the Poincaré ball, discretizes paralinguistic patterns via a hyperbolic vector-quantized prosody codebook, and captures emotion intensity through a hyperbolic emotion lens. For unsupervised adaptation, \textbf{NOVA-ARC} performs optimal-transport-based prototype alignment between source emotion prototypes and target utterances, inducing soft supervision for unlabeled speech while being stabilized through consistency regularization. Experiments show that \textbf{NOVA-ARC} delivers the strongest performance under both non-verbal-to-verbal adaptation and the complementary verbal-to-verbal transfer setting, consistently outperforming Euclidean counterparts and strong SSL baselines. To the best of our knowledge, this work is the first to move beyond verbal-speech–centric supervision by introducing a non-verbal–to–verbal transfer paradigm for SER.
\end{abstract}

\section{Introduction \& Background}

Affective understanding from voice is essential for conversational agents that aim to communicate naturally and respond appropriately to users’ emotional states \cite{10095294,10096808}. Emotional cues in the human voice arise from two complementary sources—prosodic patterns in verbal speech and non-verbal vocal bursts \cite{doi:10.1177} (e.g., laughter, sighs, cries, screams, moans)—with the latter often conveying affect largely independent of lexical content \cite{10.3389/fnins}. Non-speech vocalizations can also serve as critical communicative signals, particularly for individuals with profound disabilities or limited speech, for whom vocal sounds may remain a usable channel of expression in the absence of fluent speech \cite{MCCORMACK2010378}. Reliance on labeled verbal speech makes SER hard to scale beyond a few well-resourced languages: annotations are scarce in low-resource settings \cite{gong25b_interspeech}, and multilingual corpora differ in label conventions, speakers, and recording conditions, producing substantial domain mismatch \cite{mote25b_interspeech,mote24_interspeech}. To mitigate such mismatch, prior work has explored unsupervised domain adaptation (UDA) for SER, including adversarial alignment for cross-lingual transfer \cite{8925513ACII}, as well as approaches that reduce domain shift via KNN-based voice conversion \cite{mote24_interspeech} or vector-quantized discrete representations shared across languages \cite{mote25b_interspeech}. However, these UDA formulations still assume that emotion supervision originates from labeled verbal speech. Even so, SER supervision remains dominated by labeled speech, while non-verbal emotion recognition has only recently received broader attention \cite{10889257}. While multilingual pretraining and transfer learning improve robustness, the fundamental bottleneck persists: emotion supervision is still collected from verbal speech and is unevenly distributed across languages \cite{upadhyay24_interspeech,9747417,10889008}. Consequently, progress has largely come from improving encoders and transfer recipes, while leaving the supervision signal unchanged. A growing line of research now targets non-verbal vocalizations directly, motivated by their affective richness and the distinct modeling challenges they pose \cite{shah-johnson-2025-n}. Taken together, these observations motivate revisiting the source of emotion supervision in SER—especially in multilingual settings where labeled data is scarce. The central motivation of this study is that the observation that non-verbal vocalizations offer a stronger supervision signal for multilingual SER by disentangling affective expression from language-dependent lexical content. In verbal speech, emotion labels are inevitably entangled with words, phonotactics, and language-dependent expressive conventions; consequently, models trained on verbal emotion corpora can overfit to lexical/phonetic correlates that do not transfer across languages and domains where as Non-verbal vocalizations offer a cleaner alternative: laughter, sobs, gasps, and sighs arise from shared physiological mechanisms and are dominated by paralinguistic acoustics—voicing, spectral tilt, intensity dynamics, and temporal modulation—making their emotion supervision inherently more transferable and language-agnostic. \textit{We hypothesize that training emotion models on labeled non-verbal vocalizations—where affect is disentangled from linguistic content—learns a language-agnostic affective representation that forms a stronger and more transferable foundation for verbal emotion recognition across languages than models trained on verbal speech.} To validate this hypothesis, we conceptualize multilingual SER as unsupervised non-verbal-to-verbal transfer, where emotion supervision is learned from labeled non-verbal vocalizations and adapted to unlabeled verbal speech across languages without target-language emotion labels. In this study, we introduced \textbf{\texttt{NOVA-ARC:}} \textbf{NO}n-verbal to \textbf{V}erbal \textbf{A}daptation via hyperbolic \textbf{A}lignment, \textbf{R}adial calibration, and \textbf{C}odebook tokens, is a geometry-aware framework for unsupervised non-verbal-to-verbal emotion transfer. The model first learns affective representations from labeled non-verbal vocalizations using a self-supervised backbone, which are projected into hyperbolic space to capture hierarchical emotion structure. Prosodic patterns are discretized via a hyperbolic vector-quantized codebook and fused with continuous embeddings through a hyperbolic bottleneck. To adapt to unlabeled verbal speech across languages, \textbf{\texttt{NOVA-ARC}} aligns target utterances to source emotion prototypes using hyperbolic optimal prototype transport, inducing soft supervision without target emotion labels. Consistency regularization further stabilizes adaptation on unlabeled target speech. 

\noindent \textbf{Accordingly, this study make the following contributions:} 
\vspace{-0.3cm}
\begin{itemize}
    \item We propose a new formulation for multilingual SER: we leverage labeled non-verbal expressions to supervise training and adapt to unlabeled verbal utterances from the target domain without accessing emotion labels.
    \vspace{-0.3cm}
    \item We conduct comprehensive experiments on ASVP-ESD (non-verbal \& verbal) and five verbal SER datasets (MESD, AESDD, RAVDESS, Emo-DB, CREMA-D).
\vspace{-0.3cm}
    \item We introduced \textbf{NOVA-ARC}, a geometry-aware approach that uses hyperbolic representations to capture affective structure and a prototype-transport objective to adapt to unlabeled verbal speech, together with mechanisms for prosody tokenization and intensity calibration.
\vspace{-0.3cm}
    \item We find that \textbf{NOVA-ARC} reliably improves over competitive speech-SSL baselines and Euclidean transfer baselines across datasets and languages under an unlabeled-target setting.
\vspace{-0.3cm}
    \item To the best of our knowledge, \textbf{NOVA-ARC} is the first end-to-end framework to formulate low-resource multilingual SER as unsupervised transfer from labeled non-verbal vocalizations to unlabeled verbal speech.
    
\end{itemize}
\vspace{-0.3cm}

\blfootnote{The  project page is available at the following link: \url{https://helixometry.github.io/NOVA-ARC---ACL26/}.}




\section{Pre-Trained Representation}

We consider a set of speech foundation encoders, including voc2vec\footnote{\url{https://github.com/koudounasalkis/voc2vec}} \cite{10890672}, WavLM\footnote{\url{https://huggingface.co/microsoft/wavlm}} \cite{chen2022wavlm}, wav2vec~2.0\footnote{\url{https://huggingface.co/facebook/wav2vec2}} \cite{baevski2020wav2vec}, and MMS\footnote{\url{https://huggingface.co/facebook/mms-1b}} \cite{pratap2024scaling}. wav2vec~2.0 and WavLM are self-supervised PTMs trained with masked prediction objectives; WavLM additionally incorporates a denoising-oriented pretraining design to improve robustness and downstream generalization. MMS scales wav2vec~2.0-style self-supervised pretraining to broad multilingual coverage by training wav2vec~2.0 models for 1{,}406 languages on $\sim$491K hours of unlabeled speech, drawing data from multiple corpora. To better model non-verbal vocalizations, we additionally use voc2vec, a foundation model explicitly tailored for non-verbal human sounds, pre-trained with SSL on 10 open-source non-verbal datasets totaling $\sim$125 hours; it is built on the wav2vec~2.0 framework and follows the wav2vec~2.0 base architecture. For feature extraction, we resample all audio to 16,kHz and average-pool the final hidden-layer frame representations to obtain utterance-level embeddings. Representation dimensionalities are 768 (voc2vec, WavLM, wav2vec~2.0) and 1024 (MMS-1B). A detailed description of the pretrained models used in this study is provided in Appendix~\ref{PTMS}.

\section{Modeling}

\subsection{Downstream classifier} 
The frame-level representations extracted from each front-end model are used as input to a lightweight temporal CNN. We adopt a CNN head, widely used downstream choice in related work \cite{10889257}. The classifier consists of two convolutional blocks, where each block includes a 1D convolution followed by max-pooling along the temporal axis. The first block uses 64 filters with a kernel size of 3, and the second block uses 128 filters with the same kernel size. The resulting feature maps are then flattened and passed through a fully connected layer with 128 hidden units (ReLU). Finally, a softmax output layer produces emotion posteriors, with the number of output neurons set to the number of target emotion classes. Across the different front-end representations, the total number of trainable parameters in the CNN head ranges from 5.2M to 8.5M, depending on the embedding dimensionality of the front-end.
\begin{figure*}[]
    \centering
    \includegraphics[width=0.98\linewidth]{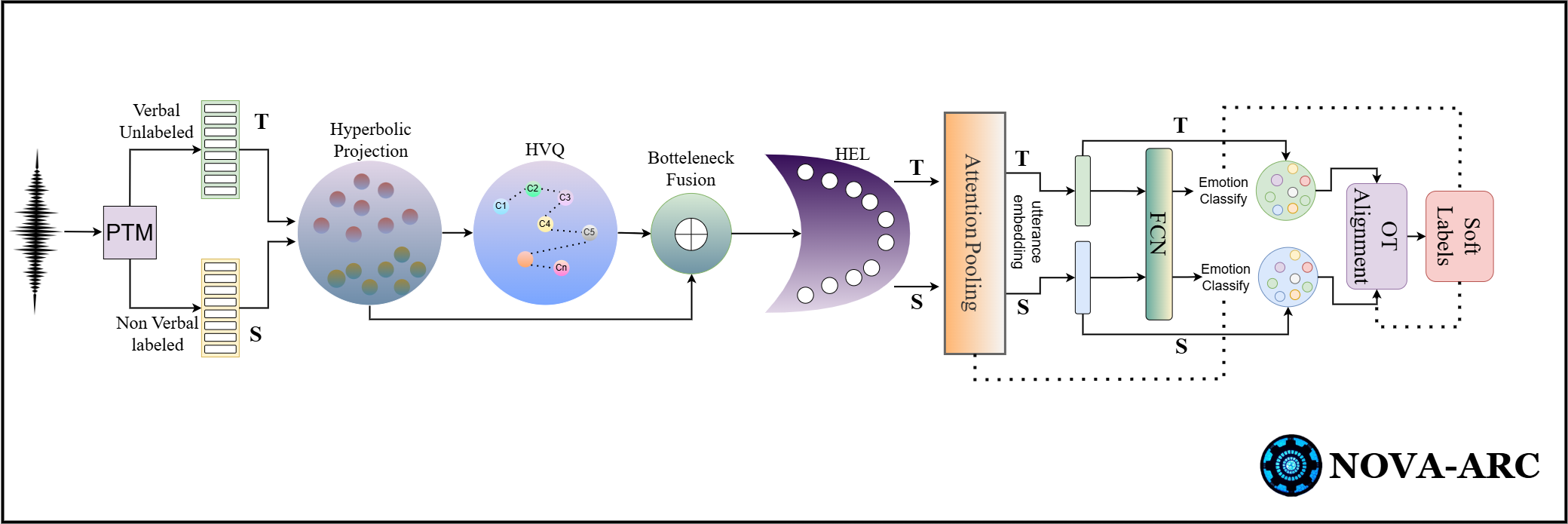}
    \caption{Proposed Framework Overview: \textbf{\texttt{NOVA-ARC}}}
    \label{archi}
\end{figure*}
\begin{table}[!hbt]
\centering
\scriptsize
\setlength{\tabcolsep}{3pt}

\begin{tabular}{l|cc|cc|cc|cc}
\toprule
\multirow{3}{*}{\textbf{Data}} &
\multicolumn{2}{c|}{\textbf{V2C}} &
\multicolumn{2}{c|}{\textbf{WVM}} &
\multicolumn{2}{c|}{\textbf{WV2}} &
\multicolumn{2}{c}{\textbf{MMS}} \\
\cmidrule(lr){2-3}\cmidrule(lr){4-7}\cmidrule(lr){8-9}
& \textbf{A \(\uparrow\)} & \textbf{F1 \(\uparrow\)} & \textbf{A \(\uparrow\)} & \textbf{F1 \(\uparrow\)} & \textbf{A \(\uparrow\)} & \textbf{F1 \(\uparrow\)} & \textbf{A \(\uparrow\)} & \textbf{F1 \(\uparrow\)} \\
\midrule

\multicolumn{9}{c}{\textbf{Verbal}} \\
\midrule
APD (V) & 32.67 &30.41  & 84.39 & 82.57 &80.56  &77.90  &87.63 &85.78  \\
MESD   &  49.02& 46.91 & 68.35 & 67.24 &72.63  & 71.98 &96.47  &93.85  \\
AESD   &  35.86& 34.19 & 84.27 &  82.94&73.41  &70.98  &61.59  &60.12  \\
RVDS   & 36.51 &33.97  & 85.69 &82.90  &76.38  &74.69  &83.67  &82.05  \\
EMDB   & 44.69 &42.28  & 96.31 &94.82  &97.67  &95.38  &94.15  & 91.82 \\
CRMD   & 42.31 &41.06  &86.71  &85.20  &86.02  &84.47  &79.23  & 76.91 \\
\midrule
\multicolumn{9}{c}{\textbf{Non Verbal}} \\
\midrule
APD (NV) & 95.26 &93.79  & 63.61 & 60.92 & 58.92 &56.47  & 46.03 &43.65  \\
\bottomrule
\end{tabular}
\caption{Cross-corpus performance. Abbreviations used : V2C = voc2vec, WVM = WavLM, WV2 = wav2vec 2.0, MMS = MMS. Datasets: APD = ASVP-ESD (V = verbal split, NV = non-verbal split), MESD = MESD, AESD = AESDD, RVDS = RAVDESS, EMDB = Emo-DB, CRMD = CREMA-D. Results are reported as accuracy (A) and macro-F1 (F1). The same abbreviations are used in Tables~\ref{tab:2} and~\ref{tab:3}.}

\label{tab:1}
\end{table}
\subsection{Proposed Framework: \textbf{\texttt{NOVA-ARC}}} 
We propose, \textbf{\texttt{NOVA-ARC}}, a hyperbolic prosody-tokenization and prototype-transport framework that leverages labeled nonverbal vocalizations to adapt SER to unlabeled verbal speech. The overall pipeline of \textbf{\texttt{NOVA-ARC}} is illustrated in Figure~\ref{archi}. We study unsupervised domain adaptation (UDA) for speech emotion recognition (SER) from labeled non-verbal vocalizations (NVV) as source to unlabeled verbal speech (UVS) as target. We define the labeled source domain as a set of NVV waveforms paired with emotion labels:
\begin{equation*}
\mathcal{D}_S=\{(x_i^S,y_i)\}_{i=1}^{N_S}, \quad y_i\in\{1,\dots,C\}
\end{equation*}
where $x_i^S$ is a NVV waveform (e.g., laugh/cry/sigh) and $y_i$ is its emotion label. We define the unlabeled target domain as a set of UVS waveforms:
\begin{equation*}
\mathcal{D}_T=\{x_j^T\}_{j=1}^{N_T}
\end{equation*}
where $x_j^T$ is a UVS waveform containing lexical content, and we do not use transcripts or emotion labels during target training. We share all model components across NVV and UVS and denote the parameters as:
\begin{equation*}
\theta=\{E,\;W_p,b_p,\;\mathcal{C},\;W_b,b_b,\;\alpha,\;w,\;W_e,b_e\}
\end{equation*}
Here, $E$ is a self-supervised acoustic encoder; $(W_p,b_p)$ projects encoder features to the latent space; $\mathcal{C}$ is a hyperbolic VQ codebook; $(W_b,b_b)$ is a bottleneck projection; $\alpha$ is the Hyperbolic Emotion Lens (HEL) parameter; $w$ is the attention pooling vector; and $(W_e,b_e)$ is the emotion classifier head. In our experiments, we instantiate $E$ with voc2vec, WavLM, wav2vec~2.0, or MMS. To make the geometry comparison as fair as possible, we construct a Euclidean counterpart of \textbf{\texttt{NOVA-ARC}} with the same overall architecture and training setup, replacing only the hyperbolic operations with Euclidean ones. The Euclidean model retains the same key components, including prototype-based transport, prosody tokenization, and intensity calibration. This allows the comparison to mainly isolate the role of representation geometry. \newline
\noindent \textbf{Hyperbolic Preliminaries:} We model embeddings in the $d$-dimensional Poincar\'e ball with curvature $-c$ ($c>0$):
\begin{equation*}
\mathbb{D}_c^d=\{\mathbf{x}\in\mathbb{R}^d: c\|\mathbf{x}\|^2<1\}
\end{equation*}
\noindent \textbf{Why hyperbolic geometry?}
We adopt hyperbolic geometry because it provides a low-distortion space for organizing representations that exhibit hierarchical structure; in our setting, this helps retain the relative organization of affective cues when moving from NVV to UVS. This is well matched to the hierarchical nature of emotions, where broad affective families can be refined into more specific categories and intensities. Recent SER work \cite{phukan25b_interspeech} shows that aligning and combining speech representations in hyperbolic space better preserves their hierarchical relationships and reduces geometric distortion, leading to more effective integration of complementary cues. Motivated by these findings, \textbf{NOVA-ARC} operates in the Poincar'e ball for prosody tokenization, continuous--discrete integration, and prototype alignment. Next, for any hyperbolic operations, we use the exponential and logarithmic maps at the origin, M\"obius addition, and the associated Poincar\'e distance, and we add a small $\varepsilon>0$ wherever needed for numerical stability. We map tangent vectors to the Poincar\'e ball using the exponential map at the origin:
\begin{equation*}
\exp_{\mathbf{0}}^c(\mathbf{v})=
\tanh(\sqrt{c}\|\mathbf{v}\|)\frac{\mathbf{v}}{\sqrt{c}\|\mathbf{v}\|+\varepsilon}
\end{equation*}
Next, we map points from the Poincar\'e ball back to the tangent space using the logarithmic map at the origin:
\begin{equation*}
\log_{\mathbf{0}}^c(\mathbf{x})=
\frac{1}{\sqrt{c}}\operatorname{arctanh}(\sqrt{c}\|\mathbf{x}\|)\frac{\mathbf{x}}{\|\mathbf{x}\|+\varepsilon}
\end{equation*}
\noindent \textbf{M\"obius addition and Poincar\'e distance.}
We combine two hyperbolic points using M\"obius addition and measure similarity using the corresponding Poincar\'e distance. For $\mathbf{x},\mathbf{y}\in\mathbb{D}_c^d$, we define
\begin{equation*}
\small
\begin{gathered}
\mathbf{x}\oplus\mathbf{y}=
\frac{(1+2c\langle \mathbf{x},\mathbf{y}\rangle + c\|\mathbf{y}\|^2)\mathbf{x}
+ (1 - c\|\mathbf{x}\|^2)\mathbf{y}}
{1 + 2c\langle \mathbf{x},\mathbf{y}\rangle + c^2\|\mathbf{x}\|^2\|\mathbf{y}\|^2}
\end{gathered}
\end{equation*}
and
\begin{equation*}
d_c(\mathbf{x},\mathbf{y})
= \frac{2}{\sqrt{c}}\operatorname{arctanh}\!\Big(\sqrt{c}\;\|-\mathbf{x}\oplus\mathbf{y}\|\Big)
\end{equation*}
\noindent \textbf{Shared Forward Pass}
Given an utterance $x$ (either $x^S$ from NVV or $x^T$ from UVS), we compute frame-level hyperbolic features and an utterance-level embedding using a shared network.
We first pass $x$ through the encoder $E$ to obtain frame-wise features $\{\mathbf{z}_t\}_{t=1}^{T}$ that summarize local acoustic patterns over time. We then project these features and map them into the Poincar\'e ball, producing hyperbolic frame embeddings $\{\mathbf{x}_t\}_{t=1}^{T}$:
\begin{equation*}
\mathbf{z}_t = E(x)_t,\qquad
\mathbf{x}_t = \exp_{\mathbf{0}}^c(W_p\mathbf{z}_t+b_p)
\end{equation*}
Next, we perform prosody tokenization with a hyperbolic VQ codebook $\mathcal{C}$: each hyperbolic frame $\mathbf{x}_t$ is assigned to its nearest codeword under the Poincar\'e distance, yielding a discrete hyperbolic token $\mathbf{q}_t$. We optimize this module with the standard codebook and commitment terms (stop-gradient), ensuring the discrete tokens remain close to the continuous hyperbolic frames while preserving stable assignments. We then fuse the continuous and discrete cues directly in hyperbolic space using M\"obius addition, and compress the fused representation through a bottleneck. Concretely, we combine $\mathbf{x}_t$ and $\mathbf{q}_t$ in the Poincar\'e ball, map the result to the tangent space at the origin, apply a linear bottleneck, and map back to hyperbolic space. This yields bottlenecked hyperbolic frame embeddings $\{\mathbf{b}_t\}_{t=1}^{T}\in\mathbb{D}_c^{d_b}$. To reduce intensity mismatch between NVV and UVS, we apply the Hyperbolic Emotion Lens (HEL) as a learnable radial calibration. We log-map each $\mathbf{b}_t$ to the tangent space, decompose it into radius and direction, apply a power-law warp controlled by $\alpha$, and map it back to the Poincar\'e ball, producing calibrated frames $\{\tilde{\mathbf{b}}_t\}_{t=1}^{T}$. Finally, we form an utterance-level embedding by attention pooling in the tangent space. We log-map the calibrated frames, compute attention weights with a shared vector $w$, and take the weighted sum to obtain the utterance representation $\mathbf{u}^{\flat}$. We also keep its hyperbolic counterpart $\tilde{\mathbf{b}}=\exp_{\mathbf{0}}^c(\mathbf{u}^{\flat})$ for prototype matching and transport. Emotion prediction is produced from $\mathbf{u}^{\flat}$ using a linear classifier and softmax:
\begin{equation*}
p_{\theta}(y\mid x)=\mathrm{softmax}(W_e\mathbf{u}^{\flat}+b_e)
\end{equation*}
\noindent \textbf{Hyperbolic Emotion Prototypes} With utterance embeddings available for the labeled NVV source, we summarize each emotion class by a prototype in the same hyperbolic space. Let $\tilde{\mathbf{b}}_i^S\in\mathbb{D}_c^{d_b}$ denote the utterance-level embedding of the $i$-th source sample $x_i^S$. For each class $c\in\{1,\dots,C\}$, we compute its hyperbolic prototype as the Fr\'echet mean that minimizes the sum of squared Poincar\'e distances to all source embeddings in that class:
\begin{equation*}
\mu^{(c)}=
\arg\min_{\mathbf{z}\in\mathbb{D}_c^{d_b}}
\sum_{i:\,y_i=c} d_c^2(\mathbf{z},\tilde{\mathbf{b}}_i^S),
\quad
\mu^{(c)}\in\mathbb{D}_c^{d_b}
\end{equation*}
We collect all class prototypes into $\mathcal{M}=\{\mu^{(1)},\dots,\mu^{(C)}\}$, which serves as a compact hyperbolic representation of the source emotion structure and is later used to guide adaptation on UVS.
\noindent \textbf{Target Adaptation via Hyperbolic Optimal Prototype Transport}
After forming the source prototype set $\mathcal{M}$, we adapt to unlabeled UVS by aligning target utterances to these prototypes in hyperbolic space. For a target minibatch $\mathcal{B}_T=\{x_j^T\}_{j=1}^{n}$, we compute utterance embeddings $\tilde{\mathbf{b}}_j^T\in\mathbb{D}_c^{d_b}$ using the same shared forward pass. We then define a prototype-to-target cost based on squared Poincar\'e distance:
\begin{equation*}
M_{cj}=d_c^2\!\left(\mu^{(c)},\tilde{\mathbf{b}}_j^T\right)\in\mathbb{R}^{C\times n}
\end{equation*}
We solve an entropically regularized optimal transport problem to obtain a soft alignment between prototypes and target samples. We set the target marginal to be uniform, reflecting no prior knowledge of target labels, and we set the prototype marginal to the source class prior, which preserves the emotion mass distribution observed in NVV:
\begin{equation*}
\begin{aligned}
\mathbf{b}&=\left(\tfrac{1}{n},\dots,\tfrac{1}{n}\right)\in\Delta^n,
\qquad
a_c=\frac{N_c}{\sum_{k=1}^{C}N_k},\\
\mathbf{a}&=(a_1,\dots,a_C)\in\Delta^C
\end{aligned}
\end{equation*}
We compute the transport plan $\Pi^{\ast}\in\mathbb{R}_+^{C\times n}$ with Sinkhorn iterations:
\begin{equation*}
\small
\begin{gathered}
\Pi^{\ast} =
\arg\min_{\Pi\ge0}\ \langle \Pi,M\rangle
+\varepsilon\sum_{c=1}^{C}\sum_{j=1}^{n}\Pi_{cj}(\log\Pi_{cj}-1) \\
\text{s.t.}\ \Pi\mathbf{1}_n=\mathbf{a},\quad
\Pi^\top\mathbf{1}_C=\mathbf{b}
\end{gathered}
\end{equation*}
The resulting transport plan induces soft pseudo-labels for target samples. Since $\sum_{c}\Pi^{\ast}_{cj}=b_j=1/n$, we define
\begin{equation*}
q_{cj}=\frac{\Pi^{\ast}_{cj}}{b_j}=n\,\Pi^{\ast}_{cj},
\qquad
\sum_{c=1}^{C}q_{cj}=1
\end{equation*}
We supervise target adaptation using two complementary objectives. The first encourages geometric alignment by minimizing the transport cost in hyperbolic space:
\begin{equation*}
\small
\begin{gathered}
L_{\mathrm{OPT}}(\mathcal{B}_T)=\langle \Pi^{\ast},M\rangle
=\sum_{c=1}^{C}\sum_{j=1}^{n}\Pi^{\ast}_{cj}\,
d_c^2\!\left(\mu^{(c)},\tilde{\mathbf{b}}_j^T\right)
\end{gathered}
\end{equation*}
The second uses the OT-induced soft labels to train the classifier on target samples via a soft cross-entropy loss:
\begin{equation*}
\small
\begin{gathered}
L_{\mathrm{OT\text{-}CE}}(\mathcal{B}_T)=
-\frac{1}{n}\sum_{j=1}^{n}\sum_{c=1}^{C} q_{cj}\log p_{\theta}(y=c\mid x_j^T)
\end{gathered}
\end{equation*}
Together, these two terms align UVS utterances to the NVV prototype structure while keeping predictions consistent with the transported emotion mass.
\noindent \textbf{Overall Objective and Inference}
At each iteration, we jointly learn from labeled NVV and unlabeled UVS. Concretely, we sample a labeled source minibatch $\mathcal{B}_S$ and an unlabeled target minibatch $\mathcal{B}_T$, and minimize a unified objective that combines source supervision with target alignment:
\begin{equation*}
\begin{aligned}
\mathcal{L}(\mathcal{B}_S,\mathcal{B}_T)
&=
L_S(\mathcal{B}_S)
+
\lambda_{\mathrm{OPT}}L_{\mathrm{OPT}}(\mathcal{B}_T) \\
&\quad+
\lambda_{\mathrm{OT}}L_{\mathrm{OT\text{-}CE}}(\mathcal{B}_T)
\end{aligned}
\end{equation*}
Here, $L_S$ trains the model to recognize emotions on NVV, while the two target terms encourage UVS utterances to follow the same prototype structure and produce consistent soft-label predictions. At inference time, we use the same forward pass and output the most likely emotion class:
\begin{equation*}
\hat{y}=\arg\max_{c\in\{1,\dots,C\}} p_{\theta}(y=c\mid x^{\mathrm{test}})
\end{equation*}
\noindent \textbf{Training Details and Hyperparameters}
We train \textbf{NOVA-ARC} using AdamW for 30 epochs with cosine decay and a 10\% warmup. We use a batch size of 16 for both source and target minibatches, optimize the encoder with a learning rate of $3\times10^{-5}$ and newly added layers with $1\times10^{-4}$, apply weight decay of 0.01, and clip gradients at 1.0. We set the hyperbolic curvature to $\kappa=-1.0$ with latent dimension $d=256$ and bottleneck dimension $d_b=128$, use a VQ codebook of size $K=256$ with $\beta=0.25$ and $\lambda_{\mathrm{VQ}}=1.0$, and apply entropic OT with $\varepsilon_{\mathrm{OT}}=0.05$ for 50 Sinkhorn iterations, with $\lambda_{\mathrm{OPT}}=\lambda_{\mathrm{OT}}=1.0$. We refresh prototypes once per epoch. Detailed information is provided in Appendix~\ref{Training Details and Hyperparameters}, summarized in Table~\ref{tab:nova_arc_hparams}

\section{Experiment \& Result}

\subsection{Benchmark Dataset}
In this work, we evaluate across diverse corpora spanning multiple languages, recording conditions, speaker demographics, and speaking styles, providing a challenging testbed for nonverbal-to-verbal emotion transfer. \newline
\noindent \textbf{ASVP-ESD} \cite{landry2020asvp}: It is a realistic emotional audio corpus that includes both non-speech vocalizations and speech utterances. In our setting, we use the non-speech subset as labeled source supervision and treat the speech subset as unlabeled target data. \newline
\noindent \textbf{MESD} \cite{9629934}: The dataset includes three voice categories (female adult, male adult, and child) and was recorded in a professional studio environment. In our setting, dataset is used as a verbal target corpus for evaluating nonverbal-to-verbal emotion transfer. \newline
\noindent \textbf{AESDD} \cite{AESDD}: It is an acted speech corpus in Greek. It contains approximately $500$ utterances. We use it as unlabeled verbal target data in our transfer setting.  \newline
\noindent \textbf{RAVDESS} \cite{10.1371/journal.pone.0196391}: It is a validated emotional expression dataset comprising $7356$ recordings. We use the audio-only speech subset as an unlabeled verbal target dataset. \newline
\noindent \textbf{Emo-DB} \cite{burkhardt05b_interspeech}: It is an acted German emotional speech corpus with $800$ utterances from $10$ speakers across seven emotions; we use it as unlabeled verbal target data. \newline 
\noindent \textbf{CREMA-D} \cite{6849440}: It is a crowd-sourced multimodal acted emotion dataset. The corpus includes $12$ emotionally neutral sentences. We use the audio-only modality as verbal target-domain speech (labels reserved for evaluation).  \textit{To ensure a fair and consistent evaluation, we standardize all datasets to a shared class label space: \texttt{happy}, \texttt{anger}, \texttt{disgust}, \texttt{sadness}, \texttt{fear}.} \par
\noindent \textit{We consider the subset of emotion categories that can be matched most reliably across all datasets and standardize all corpora to a shared five-class space: happy, anger, disgust, sadness, and fear. The resulting protocol provides a cleaner basis for evaluating transfer across languages and corpora.}

\begin{table}[!hbt]
\centering
\scriptsize
\setlength{\tabcolsep}{3pt}
\renewcommand{\arraystretch}{1.15}
\begin{tabular}{l|cc|cc|cc|cc}
\toprule
\multirow{3}{*}{\textbf{Data}} &
\multicolumn{2}{c|}{\textbf{V2C}} &
\multicolumn{2}{c|}{\textbf{WVM}} &
\multicolumn{2}{c|}{\textbf{WV2}} &
\multicolumn{2}{c}{\textbf{MMS}} \\
\cmidrule(lr){2-3}\cmidrule(lr){4-5}\cmidrule(lr){6-7}\cmidrule(lr){8-9}
& \textbf{A \(\uparrow\)} & \textbf{F1 \(\uparrow\)} & \textbf{A \(\uparrow\)} & \textbf{F1 \(\uparrow\)} & \textbf{A \(\uparrow\)} & \textbf{F1 \(\uparrow\)} & \textbf{A \(\uparrow\)} & \textbf{F1 \(\uparrow\)} \\
\midrule

\multicolumn{9}{c}{\textbf{APD NV}} \\
\midrule
APD (V)        & 62.23 & 60.87 & 43.65 & 42.26 & 42.79 & 39.14 & 39.48 & 37.61 \\
MESD           & 54.71 & 51.90 & 40.13 & 38.91 & 45.36 & 44.02 & 41.62 & 38.97 \\
AESD           & 56.86 & 55.12 & 39.34 & 36.71 & 41.23 & 39.58 & 43.65 & 42.39 \\
RVDS           & 60.01 & 58.42 & 46.79 & 43.90 & 41.38 & 39.72 & 38.58 & 35.87 \\
EMDB           & 57.93 & 55.16 & 45.08 & 42.51 & 43.75 & 42.29 & 41.63 & 39.67 \\
CRMD           & 61.27 & 59.46 & 39.62 & 36.91 & 36.78 & 35.11 & 30.91 & 28.69 \\

\midrule
\multicolumn{9}{c}{\textbf{APD V}} \\
\midrule
MESD           & 30.87 & 28.41 & 25.62 & 24.19 & 26.53 & 23.98& 21.76 & 20.34 \\
AESD           & 26.09 & 23.92 & 20.47 &18.63 & 23.71 & 21.05 & 16.86 & 15.43 \\
RVDS           & 33.46 & 31.78 & 14.89 & 13.05 & 18.31 & 15.92 & 13.21 & 11.67 \\
EMDB           & 29.78 & 28.14 & 19.66 & 17.31 & 23.92 & 22.09 & 14.11 & 11.24 \\
CRMD           & 36.12 & 34.78 & 12.03 &9.41 & 17.86 & 16.21& 10.64 &7.98 \\
\bottomrule
\end{tabular}
\caption{Zero-shot cross-corpus. APD NV as the labeled source and evaluates on verbal targets (APD(V), MESD, AESD, RVDS, EMDB, CRMD). APD V as the labeled source and evaluates on the same verbal targets. Results are reported as accuracy (A) and macro-F1 (F1). Abbreviations follow Table~\ref{tab:1}.}

\label{tab:2}
\end{table}

\begin{table*}[!hbt]
\centering
\scriptsize
\setlength{\tabcolsep}{4pt}
\renewcommand{\arraystretch}{1.15}
\resizebox{\textwidth}{!}{%
\begin{tabular}{l|cc|cc|cc|cc|cc|cc|cc|cc}
\toprule
\multirow{4}{*}{\textbf{DATA}} &
\multicolumn{4}{c|}{\textbf{V2C}} &
\multicolumn{4}{c|}{\textbf{WV2}} &
\multicolumn{4}{c|}{\textbf{WVM}} &
\multicolumn{4}{c}{\textbf{MMS}} \\
\cmidrule(lr){2-5}\cmidrule(lr){6-9}\cmidrule(lr){10-13}\cmidrule(lr){14-17}
& \multicolumn{2}{c|}{\textbf{EUC}} & \multicolumn{2}{c|}{\textbf{HYP}}
& \multicolumn{2}{c|}{\textbf{EUC}} & \multicolumn{2}{c|}{\textbf{HYP}}
& \multicolumn{2}{c|}{\textbf{EUC}} & \multicolumn{2}{c|}{\textbf{HYP}}
& \multicolumn{2}{c|}{\textbf{EUC}} & \multicolumn{2}{c}{\textbf{HYP}} \\
\cmidrule(lr){2-3}\cmidrule(lr){4-5}
\cmidrule(lr){6-7}\cmidrule(lr){8-9}
\cmidrule(lr){10-11}\cmidrule(lr){12-13}
\cmidrule(lr){14-15}\cmidrule(lr){16-17}
& \textbf{A \(\uparrow\)} & \textbf{F1 \(\uparrow\)} & \textbf{A \(\uparrow\)} & \textbf{F1 \(\uparrow\)}
& \textbf{A \(\uparrow\)} & \textbf{F1 \(\uparrow\)} & \textbf{A \(\uparrow\)} & \textbf{F1 \(\uparrow\)}
& \textbf{A \(\uparrow\)} & \textbf{F1 \(\uparrow\)} & \textbf{A \(\uparrow\)} & \textbf{F1 \(\uparrow\)}
& \textbf{A \(\uparrow\)} & \textbf{F1 \(\uparrow\)} & \textbf{A \(\uparrow\)} & \textbf{F1 \(\uparrow\)} \\
\midrule

\multicolumn{17}{c}{\textbf{APD NV (Source)}} \\
\midrule
APD V   & 87.31 &85.06  &92.40  &89.79  & 81.24 & 78.91 &86.91  &84.53  & 86.51 &85.17  &91.03  &88.92  & 83.78 &81.46  & 89.43 &88.15  \\
MESD    & 84.58 &81.92  &90.67  &89.05  & 78.69 & 76.30 &84.05  &81.76  & 83.47 &80.92  &81.09  &79.36  & 81.43 &78.90  & 86.79 &83.93 \\
AESD    & 79.63 &78.21  &84.39  &82.92  & 74.61 & 71.92 &79.43  &77.21  & 78.31 &76.56  &82.98  &81.06  & 76.14 &74.67  & 82.03 &80.24 \\
RVDS    & 87.04 &85.53  &93.79  &90.61  & 81.23 & 80.41 &87.57  &85.94  & 85.19 &83.78  &92.47  &90.31  & 84.37 &83.02  & 89.51 &87.69 \\
EMDB    & 86.71 &83.69  &92.46  &90.68  & 80.11 & 77.62 &85.63  &82.73  & 86.04 &85.21  &91.26  &88.93  & 82.97 &80.56  & 88.11 &85.74 \\
CRMD    & 85.26 &84.03  &91.32  &89.87  & 79.92 & 77.04 &85.46  &83.21  & 83.94 &82.06  &90.76  &89.29  & 82.51 &79.92  & 87.94 &85.22 \\
\midrule
\multicolumn{17}{c}{\textbf{APD V (Source)}} \\
\midrule
MESD     & 75.04 & 72.91& 81.33 & 80.12 & 68.76 & 66.51 & 75.28 & 73.98 & 67.49 & 64.37 & 74.62 & 72.08 & 70.81 & 68.20 & 76.56 & 75.39 \\
AESD     & 72.46 & 71.05 & 79.19 & 76.68 & 65.92 & 63.51 & 71.22 & 69.43 & 66.45 & 63.91 & 71.03 & 70.24 & 68.39& 67.58 & 73.11 & 70.67 \\
RVDS     & 81.92 & 79.23 & 86.76 & 83.60 & 72.34 &70.58 & 77.61 & 74.98 & 73.16 & 71.81 & 78.53 &76.02& 76.42 & 73.66 & 81.39 & 79.56 \\
EMDB     & 74.31 & 71.84 & 80.59 & 79.02 & 66.15 & 63.68& 73.61 & 72.38 & 64.87 & 62.55 & 71.42 & 68.90 & 67.29 & 65.82 & 76.63 & 75.11 \\
CRMD     & 72.42 & 70.19 & 79.61 & 74.82 & 63.94 & 60.13 & 69.70 & 66.58 & 62.01 & 60.47 & 70.39 & 68.25 & 66.08 & 65.76 & 74.19 & 72.44 \\

\bottomrule
\end{tabular}%
}
\caption{Cross-corpus adaptation results of \textbf{NOVA-ARC}.}
\label{tab:3}
\end{table*}

\begin{table}[!hbt]
\centering
\small
\setlength{\tabcolsep}{14pt}
\begin{tabular}{l|cc}
\toprule
\textbf{Method} & \textbf{Acc} \(\uparrow\) & F1 \(\uparrow\) \\
\midrule
Euclidean space (E) & 87.31 & 85.06 \\
Euclidean w/o EEL & 70.01 & 46.61 \\
No VQ (cont. only) & 74.22 & 70.43 \\
Token only (disc. only) & 76.90 & 73.18 \\
Concat/MLP (no M\"obius) & 65.36 & 62.24 \\
No HEL & 72.75 & 51.44 \\
Euclidean OT & 80.24 & 75.64 \\
Adversarial DA & 53.49 & 43.76 \\
OT-UDA baseline & 50.78 & 41.33 \\
\midrule
\textbf{NOVA-ARC (full)} & \textbf{92.40} & \textbf{89.79} \\
\bottomrule
\end{tabular}
\caption{Ablation study on NVV-to-UVS transfer (APD NV as source, APD V as target).}
\label{tab:ablation}
\vspace{-0.5cm}
\end{table}

\subsection{Results and Discussion}
In this section, we present results for individual front-end models with the shared CNN head, and evaluate our proposed framework, \textbf{\texttt{NOVA-ARC}}, for non-verbal-to-verbal transfer. Table~\ref{tab:1}, examine how self-supervised acoustic representations behave under two supervision regimes: voc2vec trained with non-verbal emotion labels, and speech-oriented SSL encoders (WavLM, wav2vec~2.0, MMS) trained with verbal emotion labels. Under non-verbal supervision, voc2vec is clearly the most effective in its native setting, achieving 95.26\% accuracy on APD(NV), far ahead of the speech-pretrained alternatives. This indicates that voc2vec captures emotion cues driven more by paralinguistic patterns than by lexical or phonetic content. We also see a sharp drop for speech-oriented encoders under non-verbal supervision, pointing to a mismatch between their learned feature spaces and the acoustic structure of non-verbal vocalizations. Under the conventional SER setup, where training and evaluation are both performed on verbal emotional speech, the speech-oriented SSL encoders consistently outperform voc2vec in Table~\ref{tab:1}. This trend is expected: WavLM, wav2vec~2.0, and MMS are optimized for speech and tend to encode phonetic structure and speech-driven prosodic regularities that are helpful when emotion is conveyed through spoken utterances. In comparison, voc2vec—while highly effective for non-verbal affect—appears less tailored to modeling verbal speech content, and therefore trails the speech-SSL representations in this regime. \par
Table~\ref{tab:2} reports zero-shot transfer results with the shared CNN head under two training setups. First, we train on the non-verbal split of the Audio Set of Vocalizations with Perceived Emotions Speech Dataset (ASVP-ESD; APD NV) and evaluate on multiple verbal SER datasets: MESD, AESDD-AESD, RAVDESS-RVDS, Emo-DB-EMDB, and CREMA-D-CRMD. Second, we train on the verbal split of ASVP-ESD-APD V and evaluate on the same set of verbal datasets as a conventional reference. Under supervision from APD(NV), voc2vec consistently yields the highest zero-shot performane on all verbal target corpora. On APD(V), it reaches 62.23\%, and it remains strong on RAVDESS 60.01\%  and CREMA-D 61.27\%. In comparison, speech-oriented SSL encoders degrade substantially under the same non-verbal supervision, with accuracies typically below 47\%. This suggests that voc2vec retains emotion-related cues learned from non-verbal vocalizations that remain useful on spoken utterances, whereas speech-pretrained encoders are less compatible with the acoustic structure of non-verbal affect. Generalization between verbal corpora also turns out to be challenging. Performance drops across all encoders, although voc2vec remains comparatively stronger (36.12\% on CRMD and 33.46\% on RVDS). To check whether the gains come from the representations or simply from the transfer setting, Table~\ref{tab:2} also reports zero-shot cross-corpus verbal SER using APD(V) as the source. In this verbal-only case, speech-oriented encoders (WavLM, wav2vec~2.0, MMS) tend to rank higher than voc2vec, consistent with their stronger modeling of speech structure. Even so, performance remains well below in-domain results, showing that cross-corpus generalization is difficult even within verbal speech due to differences in recording conditions, speakers, scripts, and label distributions. Overall, the ranking flips across regimes: speech SSL is most effective for verbal-to-verbal transfer, while voc2vec is most effective when supervision comes from non-verbal vocalizations. \par
Table~\ref{tab:3} reports cross-dataset results for Euclidean and hyperbolic variants within our proposed adaptation framework, \textbf{NOVA-ARC}. For a fair comparison, the training and evaluation protocol is kept identical within each front-end, and the Euclidean and hyperbolic variants differ only in the underlying representation geometry and the associated distance operations. The gains are strongest when APD(NV) is used as the source, which matches the main goal of this work: adapting emotion supervision learned from non-verbal vocalizations to verbal emotional speech. In this setting, voc2vec with hyperbolic modeling achieves the best overall accuracy on RVDS 93.79\% and also performs strongly on APD(V) 92.40\% and EMDB 92.46\%. We also conduct an additional evaluation on APD(NV)$\rightarrow$APD(V) under a controlled 10 dB SNR condition by adding noise to the verbal target speech. In this setting, the Euclidean variant achieves 67.01\% accuracy and 62.35\% F1, while the hyperbolic variant reaches 79.44\% accuracy and 78.09\% F1. This result further strengthens the empirical picture and shows that the advantage of the hyperbolic variant is retained under moderate noise. The same trend holds for the speech-oriented encoders after fine-tuning, where the hyperbolic variant improves performance across all verbal targets, showing that the proposed adaptation is not tied to a single front-end. Using the verbal split as the source, \textbf{\texttt{NOVA-ARC}} delivers consistent gains across all verbal target corpora and across all encoders, indicating that the improvements are systematic rather than model-specific. This broad, front-end-agnostic behavior strengthens the case that \textbf{\texttt{NOVA-ARC}} contributes a genuinely transferable adaptation mechanism, not just a better feature extractor, and is especially effective in the non-verbal-supervised setting targeted in this work. 
\begin{figure*}[!hbt]
    \centering
    
    \begin{subfigure}[t]{0.24\textwidth}
        \centering
        \includegraphics[width=\linewidth]{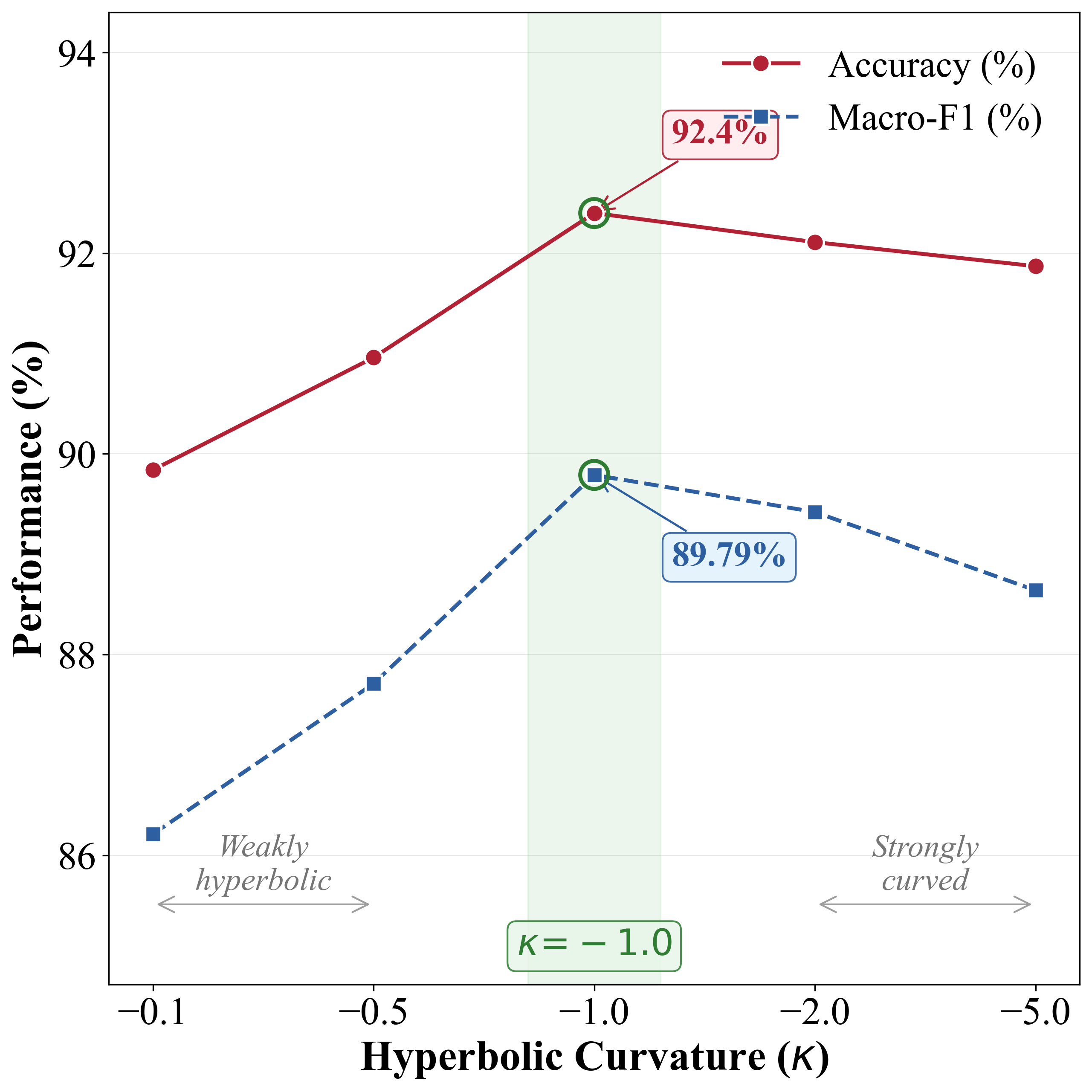}
        \label{fig:curvature_sensitivity}
    \end{subfigure}
    \hfill
    \begin{subfigure}[t]{0.24\textwidth}
        \centering
        \includegraphics[width=\linewidth]{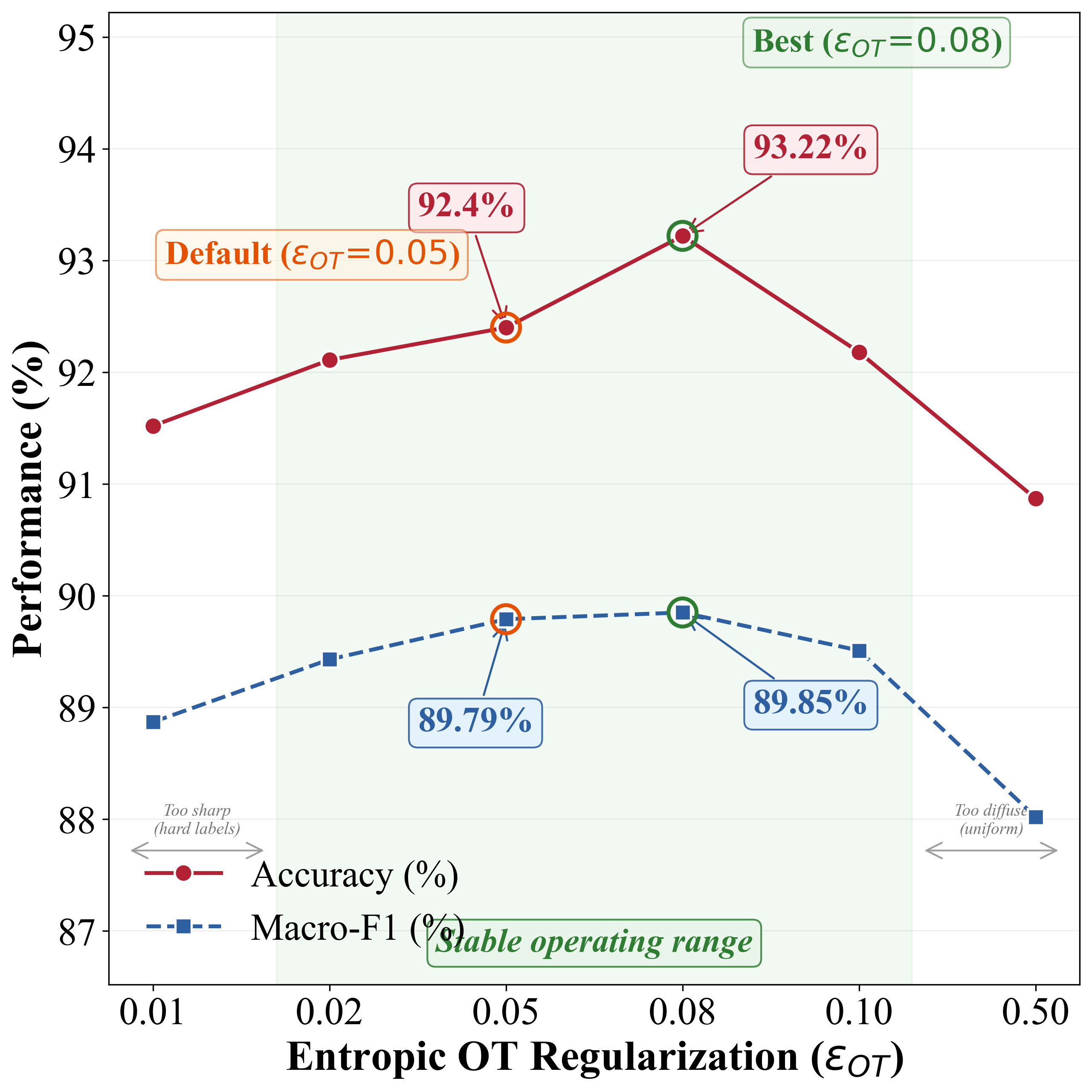}
        \label{fig:codebook_sensitivity}
    \end{subfigure}
    \hfill
    \begin{subfigure}[t]{0.24\textwidth}
        \centering
        \includegraphics[width=\linewidth]{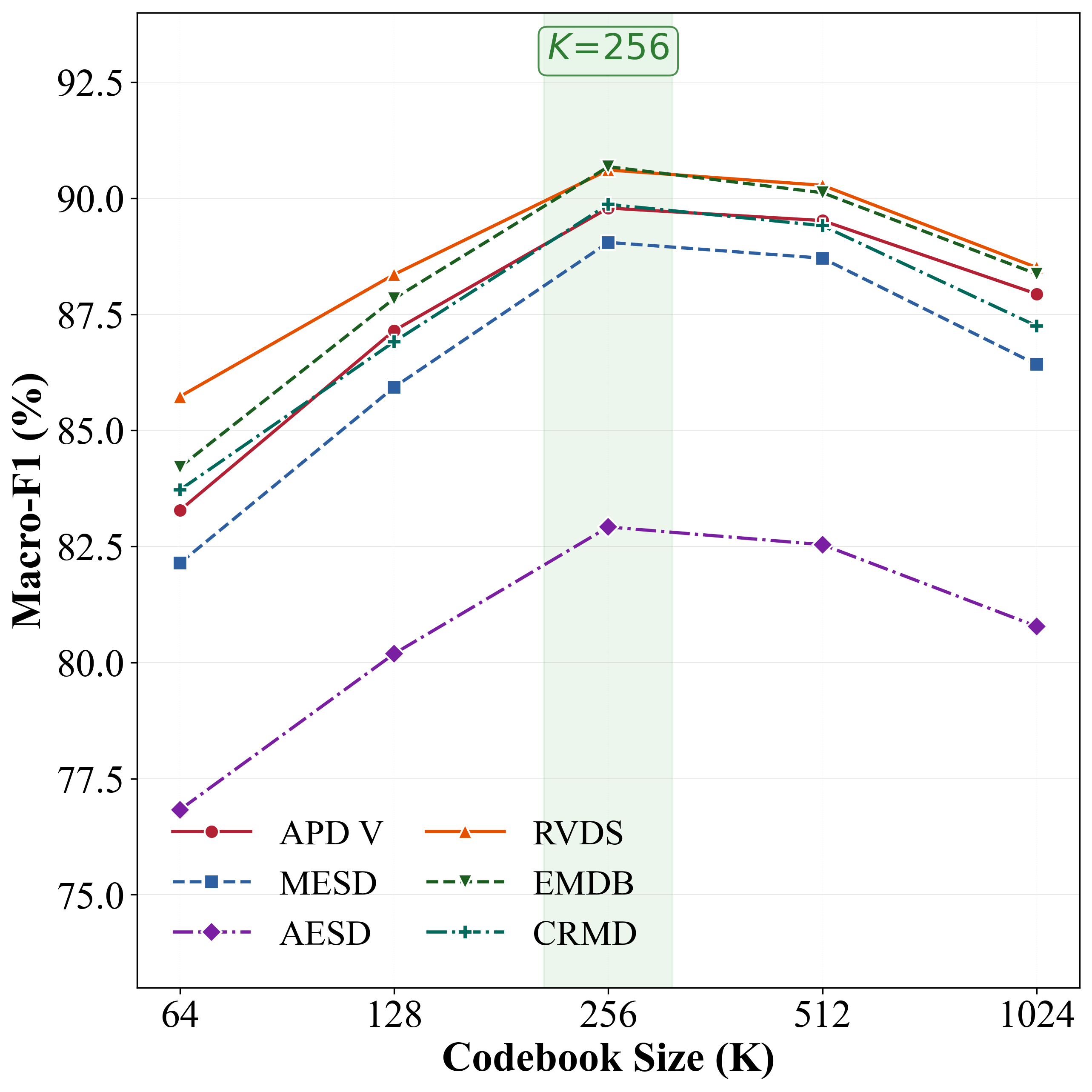}
        \label{fig:sinkhorn_sensitivity}
    \end{subfigure}
    \hfill
    \begin{subfigure}[t]{0.24\textwidth}
        \centering
        \includegraphics[width=\linewidth]{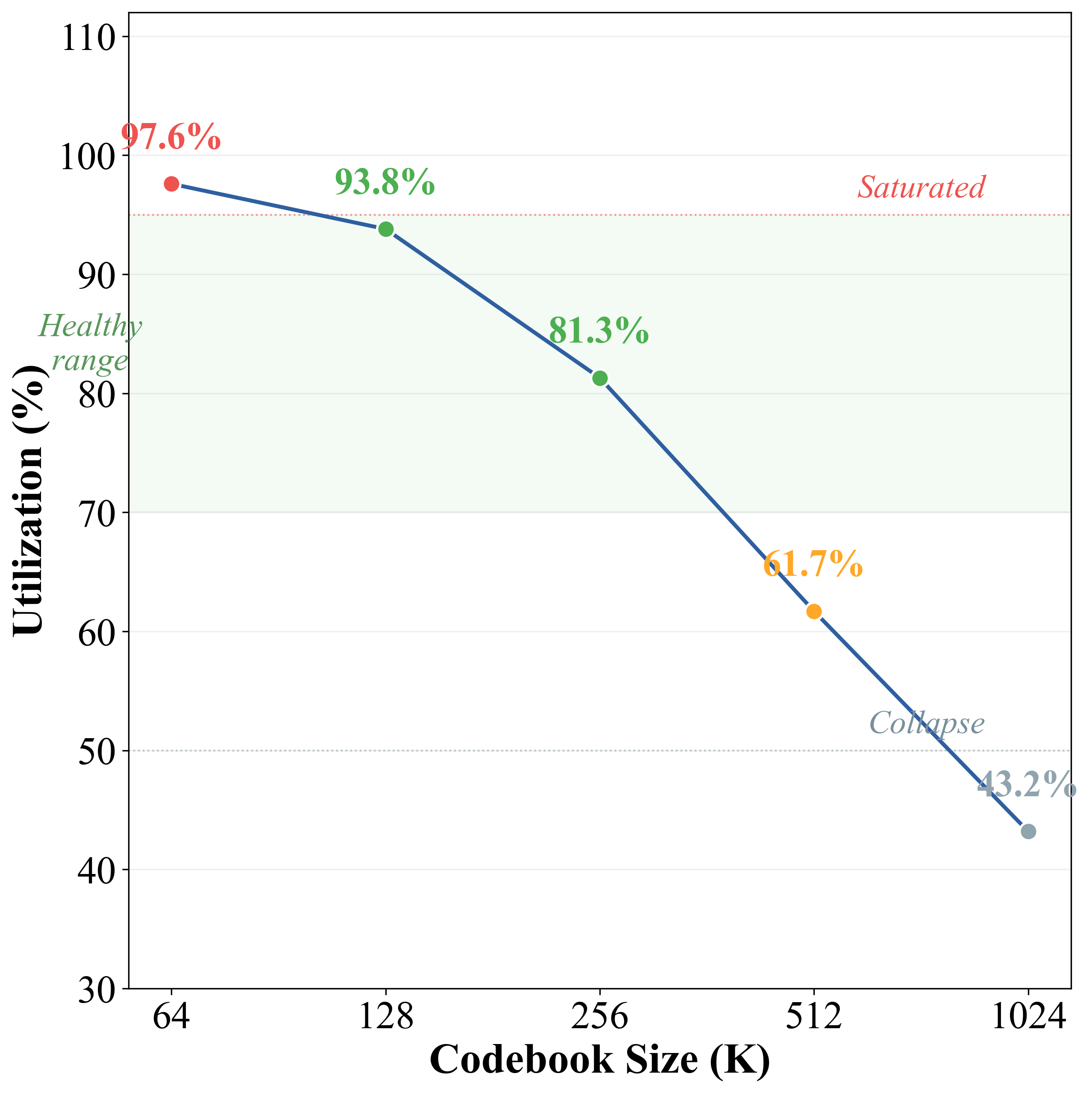}
        \label{fig:codebook_utilization}
    \end{subfigure}

 \caption{Sensitivity and codebook analysis of \textbf{NOVA-ARC} under the APD(NV)$\rightarrow$APD(V) setting, showing: (a) curvature sensitivity, (b) sensitivity to entropic OT regularization $\epsilon_{\mathrm{OT}}$, (c) codebook-size sensitivity, and (d) codebook utilization across different codebook sizes.}
    \label{fig:sensitivity_analysis}
\end{figure*}
In terms of operating behavior, \textbf{\texttt{NOVA-ARC}} is expected to work best when emotion is expressed mainly through paralinguistic cues that are shared between non-verbal vocalizations and verbal speech, such as pitch movement, energy variation, speaking rate, and voice quality. This is especially useful in settings where lexical content is weakly informative or potentially misleading, since the model is encouraged to focus on how something is spoken rather than on language-specific verbal patterns. By contrast, transfer becomes more difficult when emotion categories are close in their acoustic realization or when the correct label depends more heavily on semantic or discourse context than on prosody alone. In such cases, the model may exhibit more confusion between nearby affective states, which is consistent with the qualitative patterns observed in the appendix~\ref{sec:appendix}. For qualitative analysis, we report both t-SNE plots of the utterance embeddings and confusion matrices in the Appendix~\ref{sec:appendix}: \ref{CMAPP} and \ref{TsneAPP}, which illustrate clearer class separation and more consistent prediction patterns after adaptation. \newline
\noindent \textbf{Additional Experiment:} To further validate, we also evaluate \textbf{NOVA-ARC} on the same setup used in \cite{mote25b_interspeech}. Using their benchmark configuration, our framework attains 76.89\% accuracy and 71.43\% F1, supporting that our hyperbolic tokenization and prototype-transport design remains effective even when compared against a representative VQ-based UDA formulation.

\subsection{Ablation Study} 
\noindent Table~\ref{tab:ablation} analyzes the contribution of each component of \textbf{NOVA-ARC} for NVV-to-UVS transfer, using APD NV as the source and APD V as the target. The full model achieves 92.40\% accuracy and 89.79\% F1, and every ablation leads to a clear degradation, confirming that \textbf{NOVA-ARC} benefits from the joint design of hyperbolic modeling, prosody tokenization, intensity calibration, and transport-based alignment. Replacing the hyperbolic space with a Euclidean variant reduces performance to 87.31\%, showing that the hyperbolic geometry contributes meaningful structure for cross-domain transfer. Further ablating the Euclidean intensity calibration step reduces performance to 70.01\% accuracy and 46.61\% F1, indicating that calibration remains beneficial even within the Euclidean setting. Removing the prosody tokenization branch also hurts performance: the continuous-only variant drops to 74.22\%, and the discrete-only variant reaches 76.90\%. This indicates that continuous cues and discretized prosody tokens are complementary, and \textbf{NOVA-ARC} benefits from using both. The way these cues are integrated is also important. Replacing M\"obius integration with a tangent-space concatenation/MLP results in a large drop 65.36\%. Removing HEL decreases accuracy to 72.75\% and substantially reduces F1 to 51.44, highlighting that calibrating intensity is necessary to reduce mismatch between NVV and UVS. Finally, we ablate the adaptation mechanism. Performing OT in Euclidean space lowers performance to 80.24\%, indicating that prototype transport is more effective when distances and geometry are consistent with the hyperbolic embedding space. Standard alternatives are notably weaker: adversarial domain adaptation achieves 53.49\% and OT-based SER UDA baselines reach 50.78\%. Collectively, the ablation results show that \textbf{NOVA-ARC’s} gains come from the synergy between hyperbolic prosody tokenization, intensity-aware embedding calibration, and hyperbolic prototype transport. Figure~\ref{fig:sensitivity_analysis} complements the ablation study with a sensitivity analysis under the same APD(NV)$\rightarrow$APD(V) setting. Across curvature, entropic OT regularization, and codebook size, the selected values fall in stable operating regions rather than narrow optima. The codebook-utilization results further show that a moderate codebook size provides a good balance between representational capacity and effective usage.

\section{Conclusion}
\label{sec:conclusion}

In this study, we introduced a new perspective on low-resource multilingual SER by reframing it as unsupervised non-verbal-to-verbal transfer: learn emotion supervision from labeled non-verbal vocalizations. Through extensive evaluations on ASVP-ESD and five verbal SER benchmarks, we showed that non-verbal and verbal emotion cues induce a clear representation shift in common SSL encoders. To address this gap, we introduce \textbf{\texttt{NOVA-ARC}}, a structured adaptation approach that combines hyperbolic geometry, prosody-aware discretization, and prototype-based optimal transport for domain alignment. Across encoders and datasets, \textbf{\texttt{NOVA-ARC}} consistently improved cross-dataset generalization. Our findings establish a strong baseline for using non-verbal affect as scalable supervision for multilingual SER under realistic label constraints. \newline

\vspace{-0.3cm}
\section*{Limitations and Future Work} 

The current evaluation across multiple publicly available benchmarks demonstrates the effectiveness of proposed framework for non-verbal-to-verbal transfer in multilingual SER. A remaining direction is broader validation on spontaneous conversational speech, particularly in settings involving spontaneous dialogue and overlapping speakers that are not fully represented in the present experimental setting.

\vspace{-0.1cm}
\section*{Ethical Considerations}
This study explores low-resource multilingual SER by transferring supervision from labeled non-verbal vocalizations to unlabeled verbal speech. We use only publicly available datasets and do not collect any new data. Our framework is developed for research and benchmarking purposes. We emphasize that emotion predictions are inherently uncertain and context-dependent, and the proposed system should not be used as a stand-alone tool in high-stakes settings.

\bibliography{custom}

\appendix

\section{Appendix}
\label{sec:appendix}

\section*{Appendix}

In this appendix, we provide:
\begin{itemize}
    \item \textbf{A: Detailed Information for Pretrained Models.}
    \item \textbf{B: Training Details and Hyperparameters.} (Table~\ref{tab:nova_arc_hparams}).
    \item \textbf{C: Visualization Analysis.} Supplementary qualitative analyses through confusion matrices (Figure~\ref{fig:cm_8}) and t-SNE visualizations (Figure~\ref{fig:tsnenv}).
\end{itemize}

\subsection{Detailed Information for Pretrained Models}
\label{PTMS}

\noindent \textbf{voc2vec}\footnote{\url{https://github.com/koudounasalkis/voc2vec}} \cite{10890672}: It is a self-supervised pretrained model developed for non-verbal human vocalizations. The model follows the wav2vec 2.0 framework and is trained with self-supervised learning on 10 open-source datasets containing approximately 125 hours of non-verbal audio. The pretraining corpus includes vocalization-focused subsets from AudioSet and FreeSound, as well as datasets containing baby cries, affective dyadic interactions, communicative vocalizations, vocal imitations, and spontaneous human non-speech sounds. This pretraining setup is intended to produce representations that are better aligned with non-verbal acoustic and affective structure than conventional speech-pretrained models. \newline

\noindent \textbf{WavLM}\footnote{\url{https://huggingface.co/microsoft/wavlm}} \cite{chen2022wavlm}: It is a large-scale self-supervised pretrained speech model developed for full-stack speech processing. It extends masked speech prediction with an additional denoising objective, allowing the model to learn not only content-related information but also features useful for non-ASR speech tasks. The model further incorporates gated relative position bias in the Transformer backbone. In its large-scale setting, WavLM is pretrained on 94k hours of unlabeled speech drawn from Libri-Light, GigaSpeech, and VoxPopuli, covering diverse recording conditions and speaking styles.  \newline

\noindent \textbf{wav2vec~2.0}\footnote{\url{https://huggingface.co/facebook/wav2vec2}} \cite{baevski2020wav2vec}: It is a pretrained speech encoder built for self-supervised representation learning from raw waveform input. The model uses a convolutional feature encoder to produce latent speech representations, which are then processed by a Transformer context network. Its pretraining masks spans in the latent space and learns to identify the correct quantized target among distractors through a contrastive objective. To define these targets, wav2vec 2.0 jointly learns a quantization module that discretizes the latent representations into speech units, allowing pretraining to combine continuous contextual modeling with discrete target prediction.  \newline

\noindent \textbf{MMS}\footnote{\url{https://huggingface.co/facebook/mms-1b}} \cite{pratap2024scaling}: It is a multilingual self-supervised speech model that scales wav2vec 2.0-style pretraining to very broad language coverage. The pretrained MMS models are trained on about 491K hours of unlabeled speech spanning 1,406 languages, using data drawn from multiple corpora, including MMS-lab-U, MLS, CommonVoice, VoxLingua-107, BABEL, and VoxPopuli. This large-scale multilingual design is intended to provide broad cross-lingual speech representations, especially for languages with limited supervised resources.  \newline

\subsection{Training Details and Hyperparameters}
\label{Training Details and Hyperparameters}
Table~\ref{tab:nova_arc_hparams} lists the full set of NOVA-ARC hyperparameters used across all experiments, including the hyperbolic geometry settings, VQ configuration, optimal transport parameters, and optimization details.

\begin{table}[!hbt]
\centering
\small
\setlength{\tabcolsep}{7pt}
\begin{tabular}{l l}
\toprule
\textbf{Hyperparameter} & \textbf{Value} \\
\midrule
Hyperbolic curvature & $\kappa=-1.0$ \; \\
Hyperbolic latent dim & $d=256$ \\
Hyperbolic bottleneck dim & $d_b=128$ \\
\midrule
VQ codebook size & $K=256$ \\
VQ commitment weight & $\beta=0.25$ \\
VQ loss weight & $\lambda_{\mathrm{VQ}}=1.0$ \\
\midrule
HEL exponent (init) & $\alpha=1.0$ (learned) \\
HEL stabilizer & $\varepsilon=10^{-8}$ \\
\midrule
OT entropic reg. & $\varepsilon_{\mathrm{OT}}=0.05$ \\
Sinkhorn iterations & $L_{\mathrm{sk}}=50$ \\
OPT loss weight & $\lambda_{\mathrm{OPT}}=1.0$ \\
OT-CE loss weight & $\lambda_{\mathrm{OT}}=1.0$ \\
\midrule
Optimizer & AdamW \\
AdamW betas & (0.9, 0.98) \\
AdamW epsilon & $10^{-8}$ \\
Weight decay & 0.01 \\
LR (encoder) & $3\times10^{-5}$ \\
LR (new layers) & $1\times10^{-4}$ \\
Gradient clipping & 1.0 \\
Epochs & 30 \\
Schedule & 10\% warmup + cosine decay \\
\midrule
Batch size & $B=16$ \\
Prototype refresh & once per epoch \\
\bottomrule
\end{tabular}
\caption{NOVA-ARC hyperparameters used in all experiments.}
\label{tab:nova_arc_hparams}
\end{table}

\subsection{Visualization Analysis}
\label{infoptms}

\subsubsection{Confusion Matrices}
\label{CMAPP}
\begin{figure*}[!hbt]
    \centering
    \subfloat[]{%
        \includegraphics[width=0.22\textwidth]{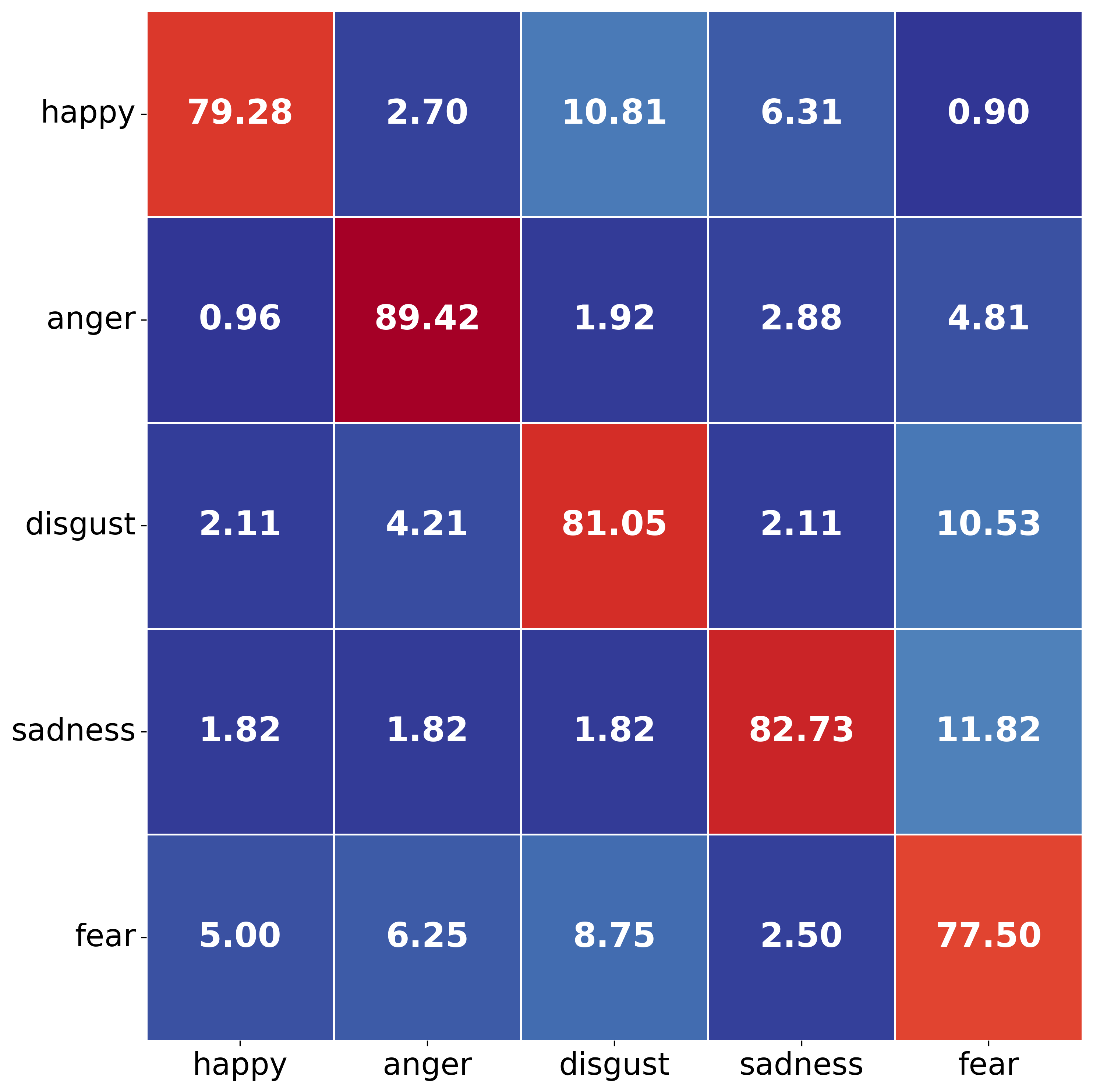}
    }
    \hspace{0.01\textwidth}
    \subfloat[]{%
        \includegraphics[width=0.22\textwidth]{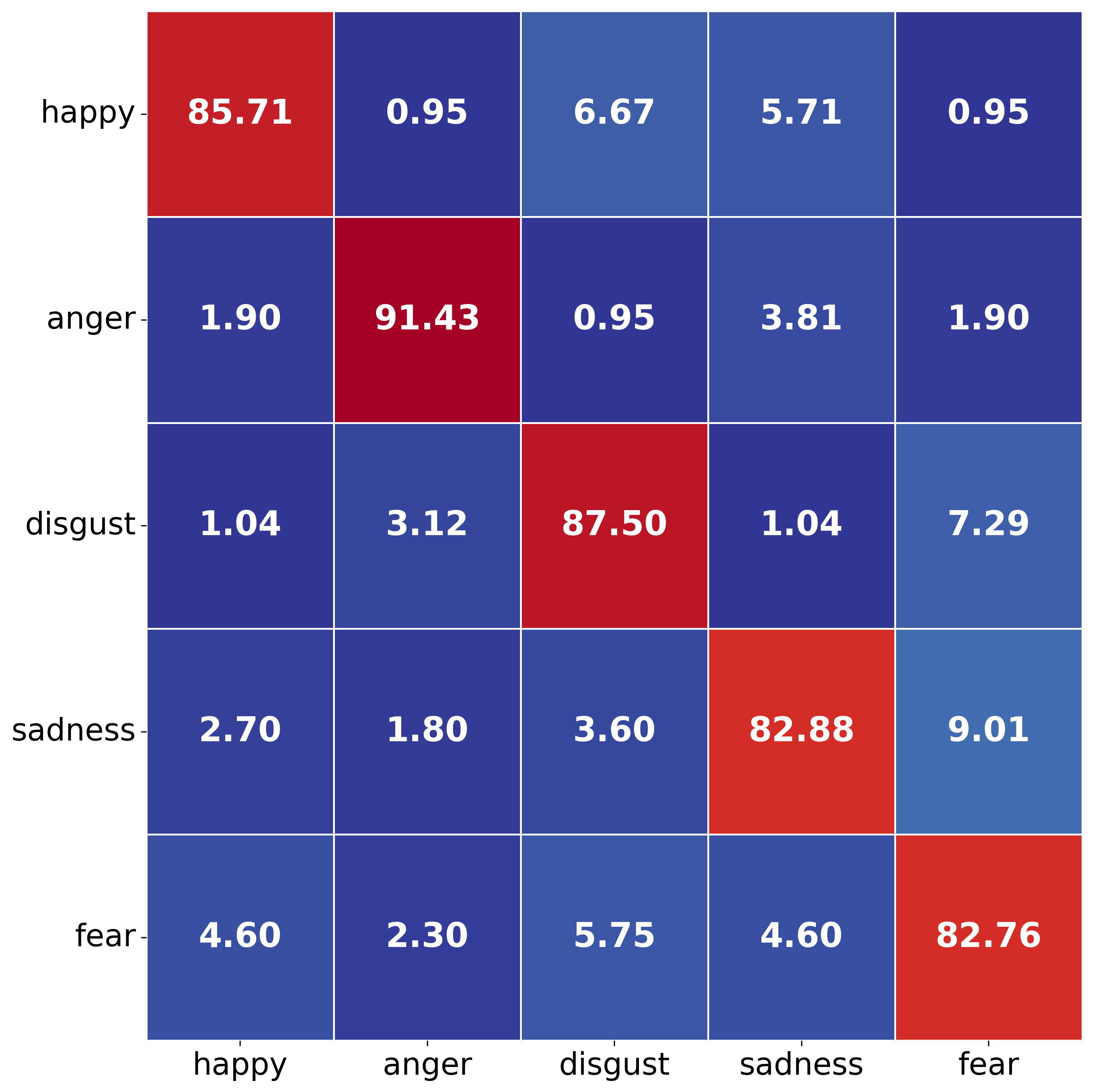}
    }
        \hspace{0.03\textwidth}
    \subfloat[]{%
        \includegraphics[width=0.22\textwidth]{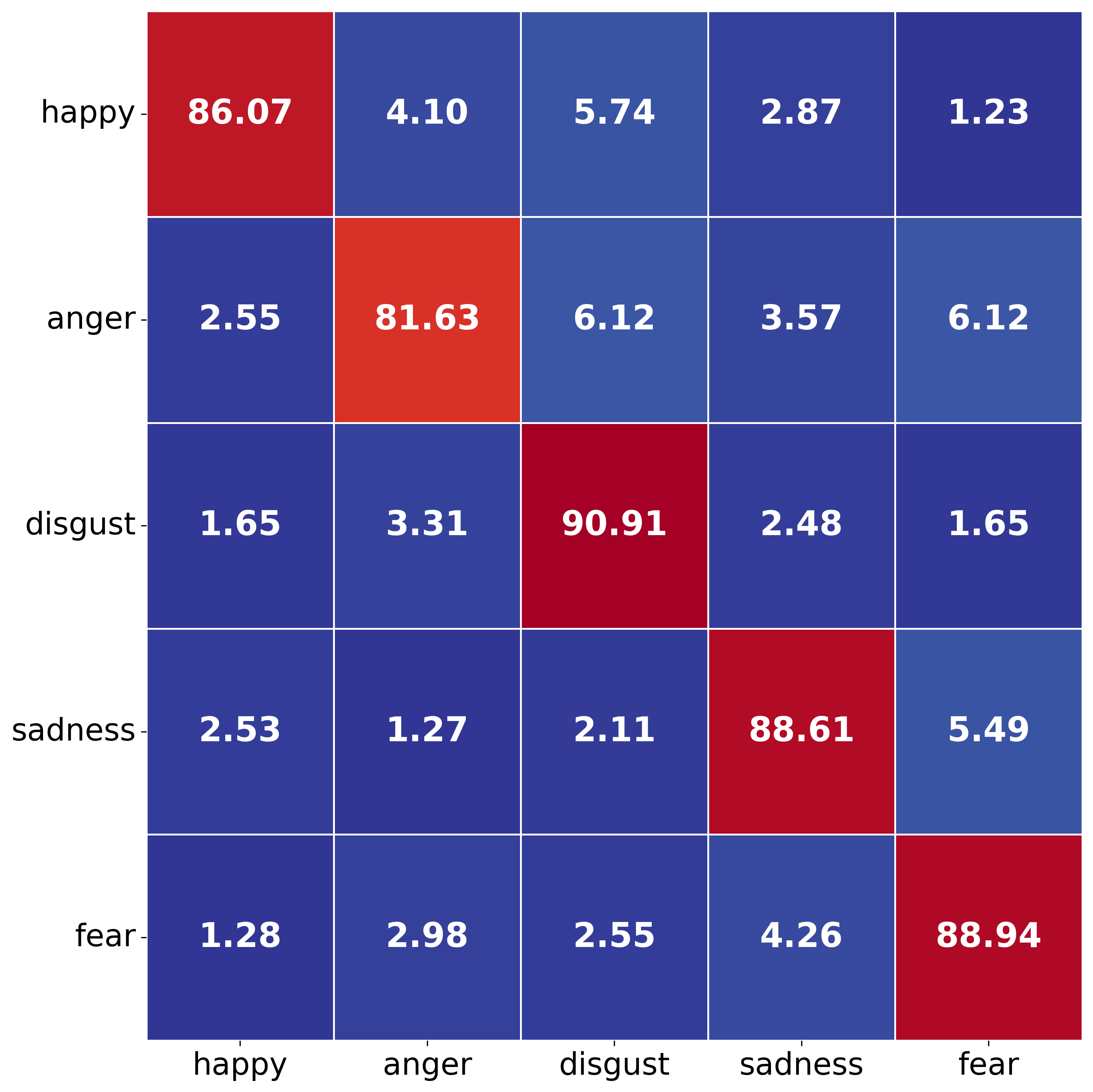}
    }
        \hspace{0.01\textwidth}
    \subfloat[]{%
        \includegraphics[width=0.22\textwidth]{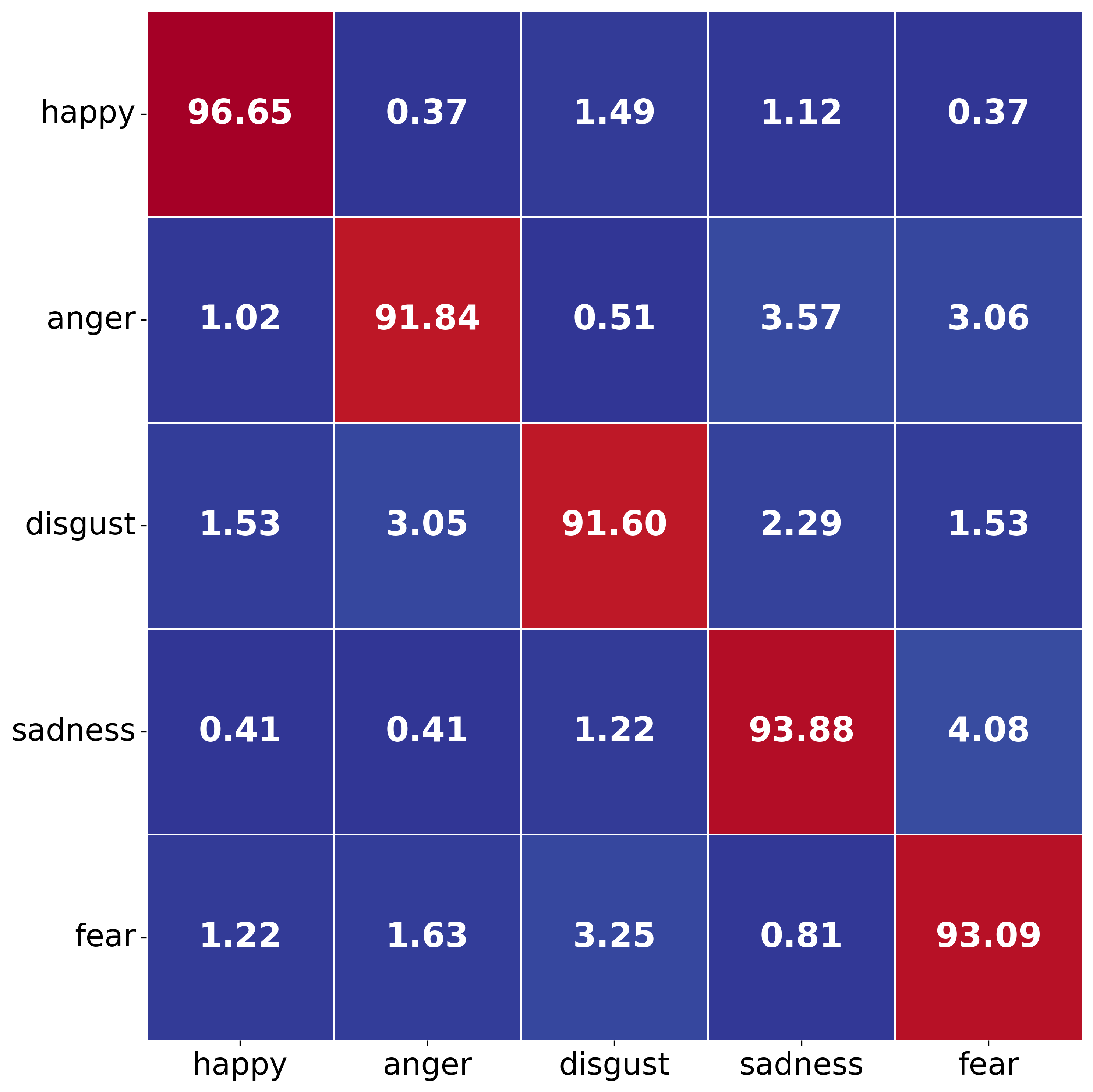}
    }
    \caption{Confusion matrices for: (a) \textbf{\texttt{NOVA-ARC}} APD-V(Source)-RAVDESS-V(Target) using Euclidean; (b) \textbf{\texttt{NOVA-ARC}} APD-V(Source)-RAVDESS-V(Target) using Hyperbolic; (c) \textbf{\texttt{NOVA-ARC}} APD-NV(Source)-RAVDESS-V(Target) using Euclidean; (d) \textbf{\texttt{NOVA-ARC}} APD-NV(Source)-RAVDESS-V(Target) using Hyperbolic. The plots provide a class-wise view of prediction reliability and the dominant error patterns under each setting.}
    \label{fig:cm_8}
\end{figure*}

Figure~\ref{fig:cm_8} reports confusion matrices for representative \textbf{NOVA-ARC} settings on ASVP-NV (source) $\rightarrow$ ASVP-V (target). These plots provide a class-wise view of prediction behavior.

\subsubsection{t-SNE Plots}
\label{TsneAPP}
\begin{figure*}[!hbt]
    \centering
    \subfloat[]{%
        \includegraphics[width=0.32\textwidth]{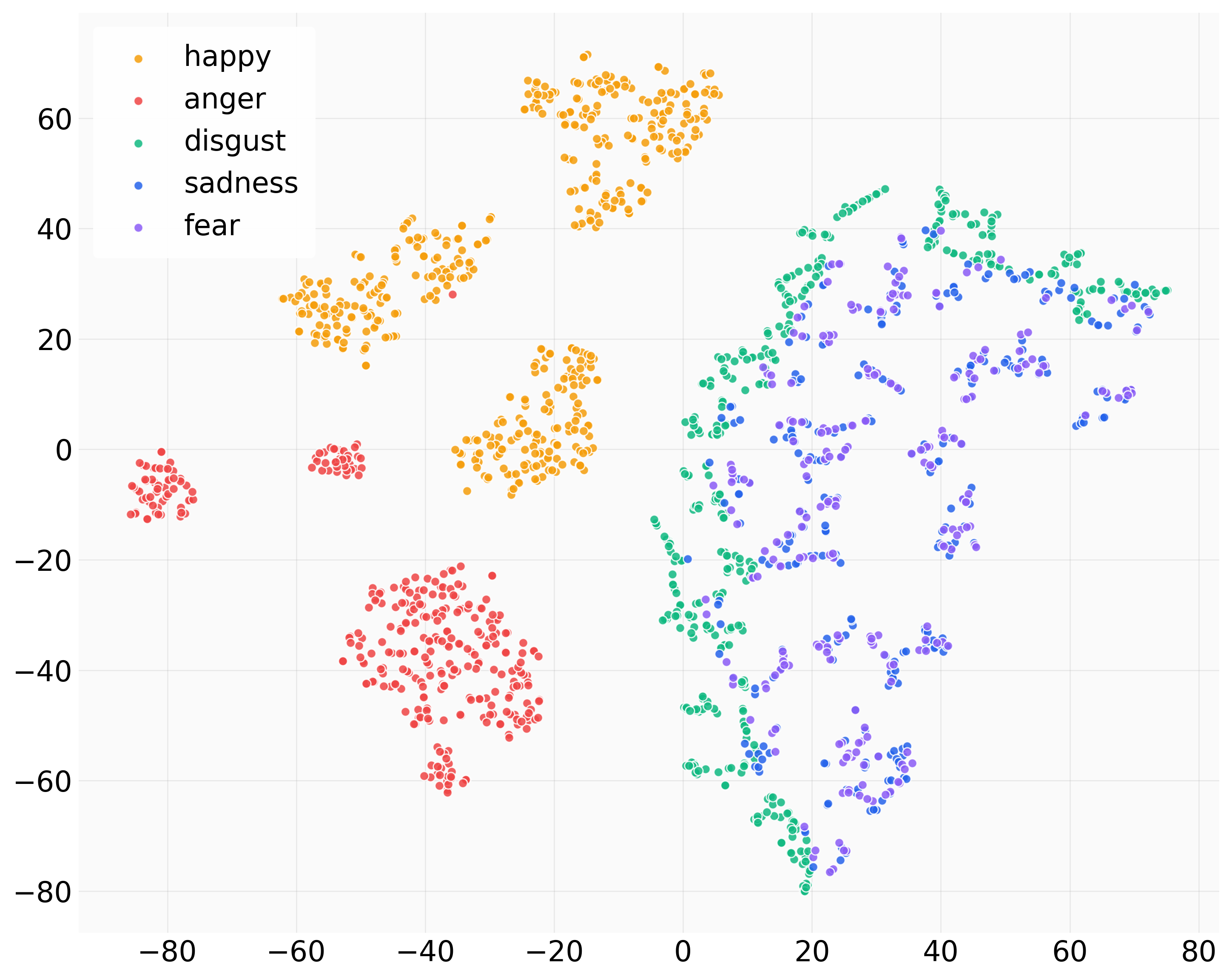}
    }
    \hspace{0.03\textwidth}
    \subfloat[]{%
        \includegraphics[width=0.32\textwidth]{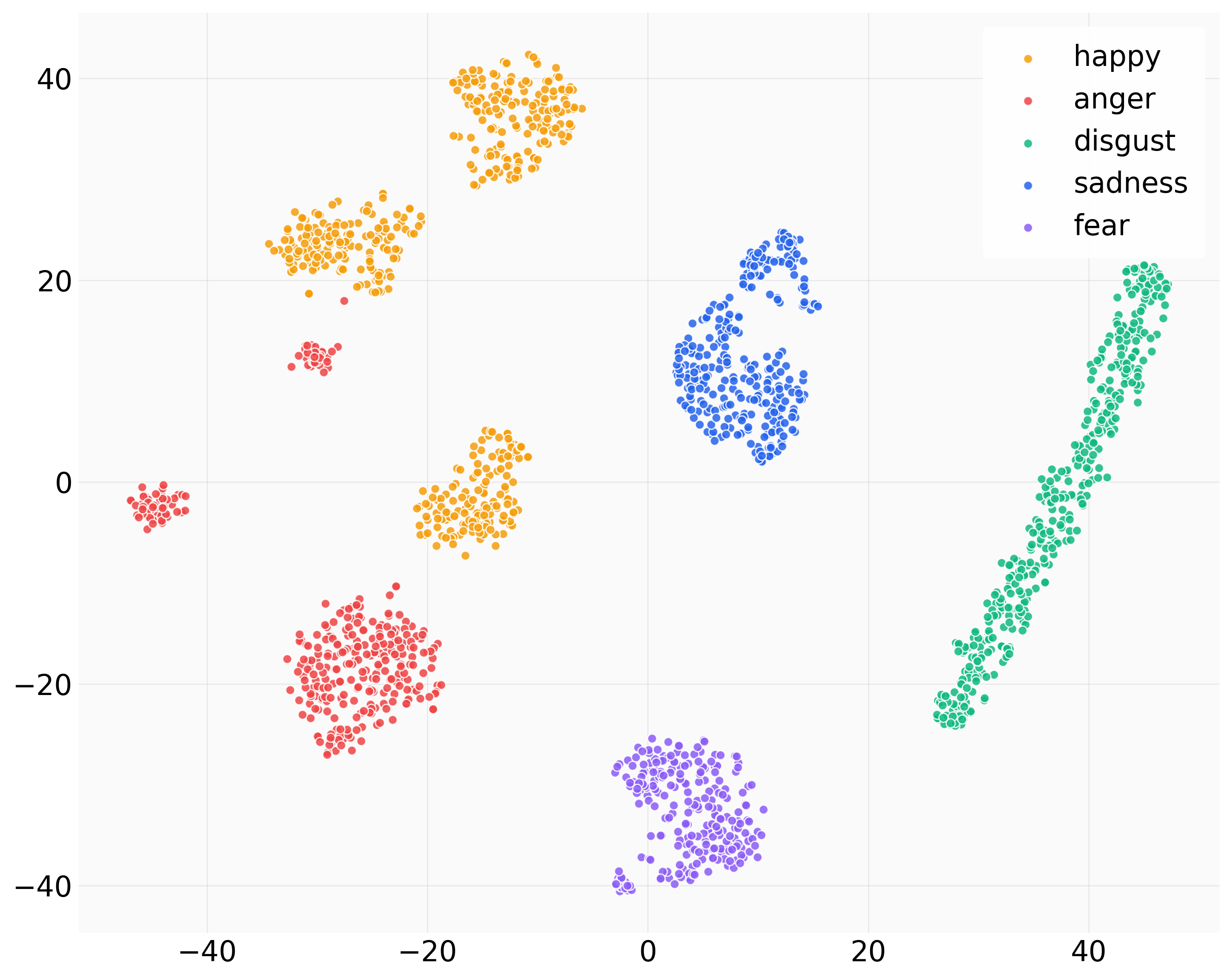}
    }\\[1ex]

    \subfloat[]{%
        \includegraphics[width=0.32\textwidth]{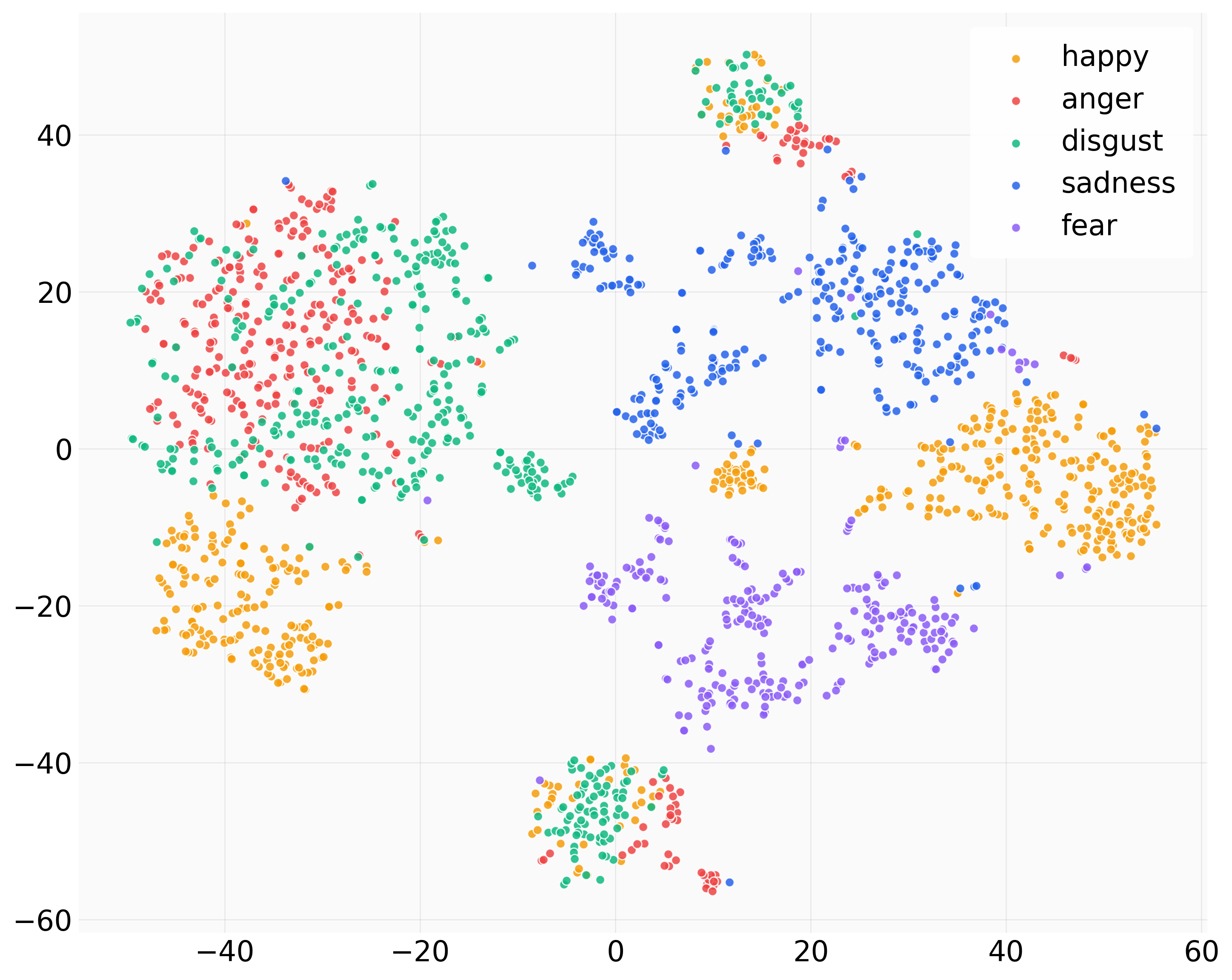}
    }
    \hspace{0.03\textwidth}
    \subfloat[]{%
        \includegraphics[width=0.32\textwidth]{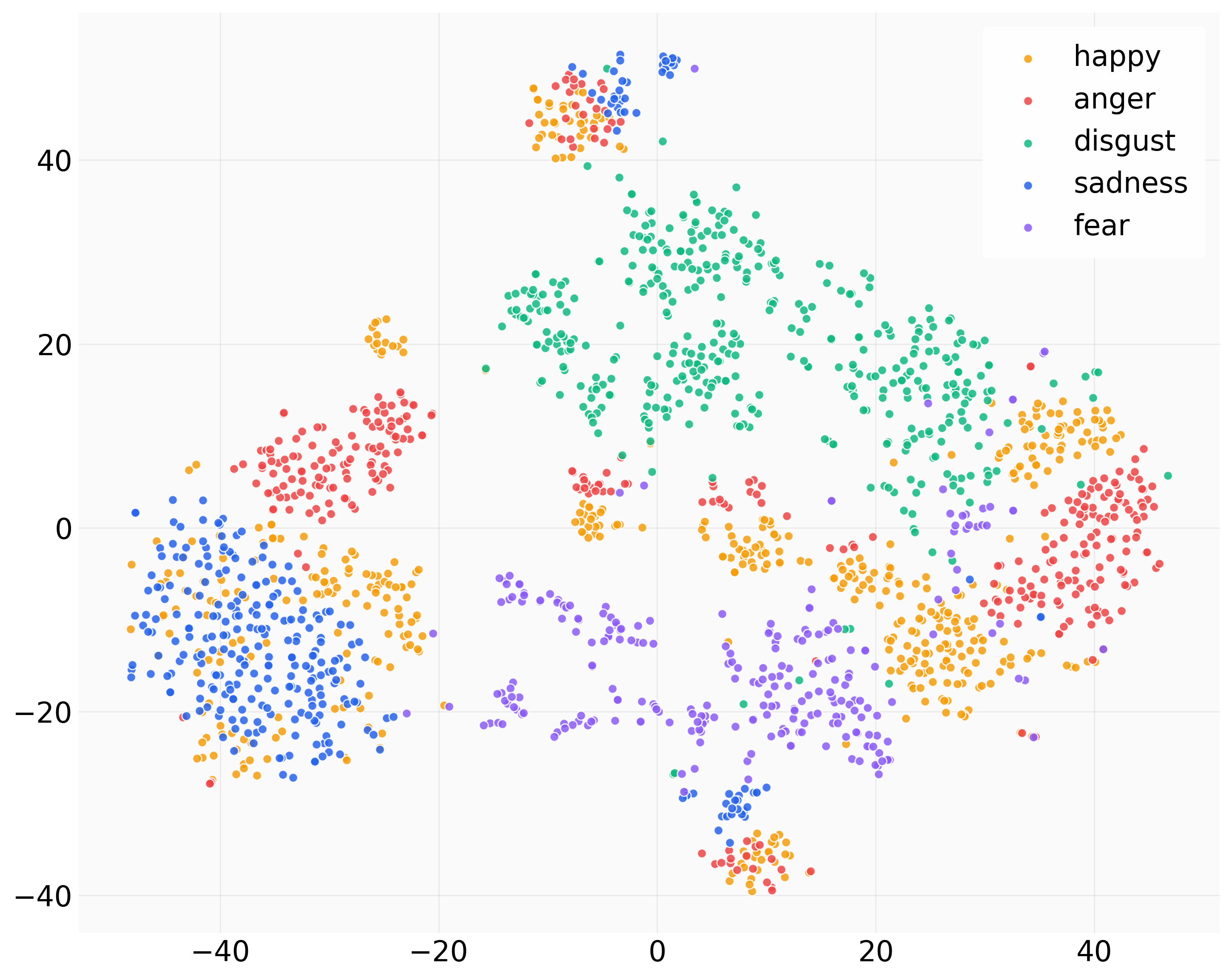}
    }\\[1ex]

    \subfloat[]{%
        \includegraphics[width=0.32\textwidth]{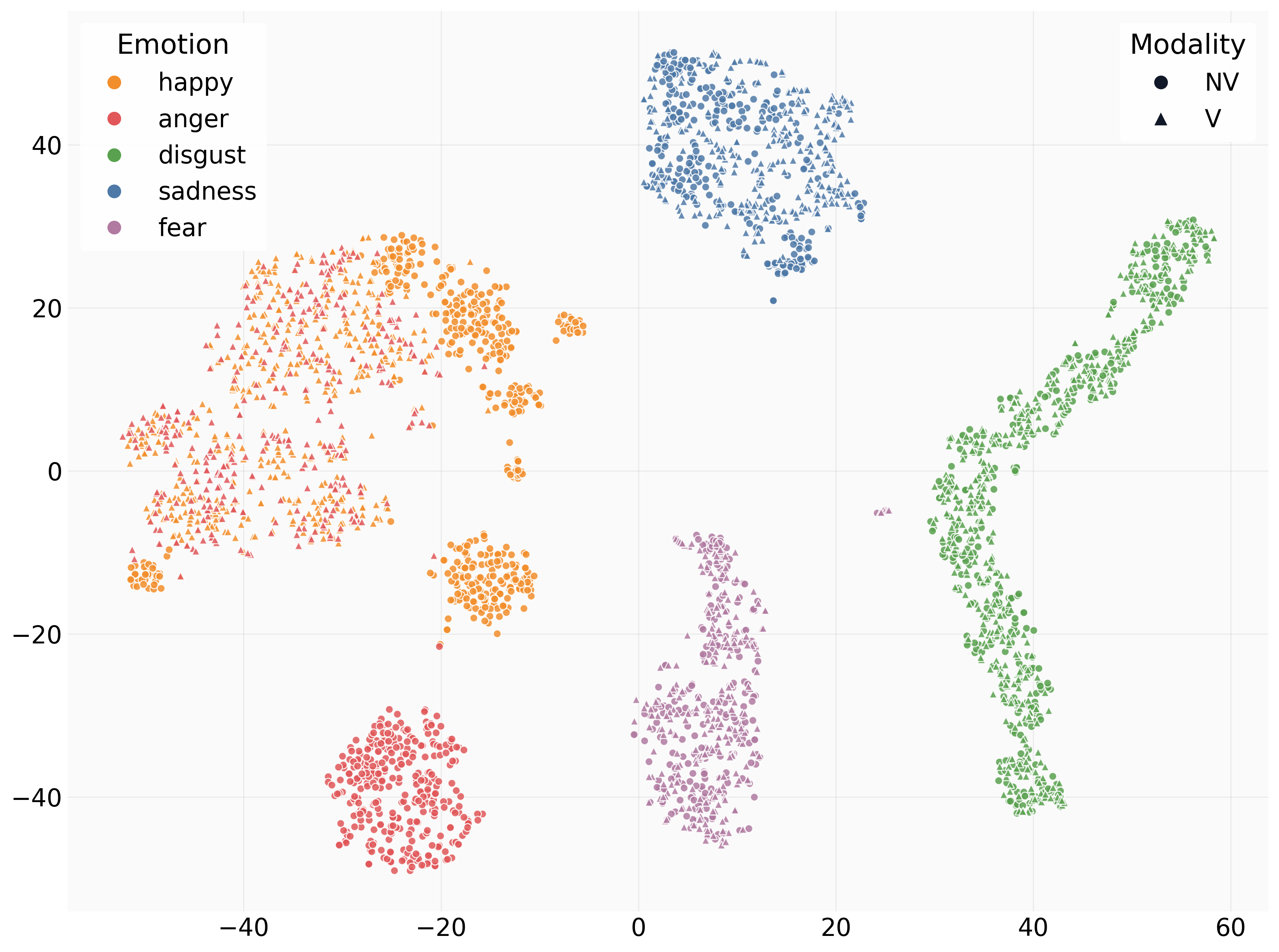}
    }
    \hspace{0.03\textwidth}
    \subfloat[]{%
        \includegraphics[width=0.32\textwidth]{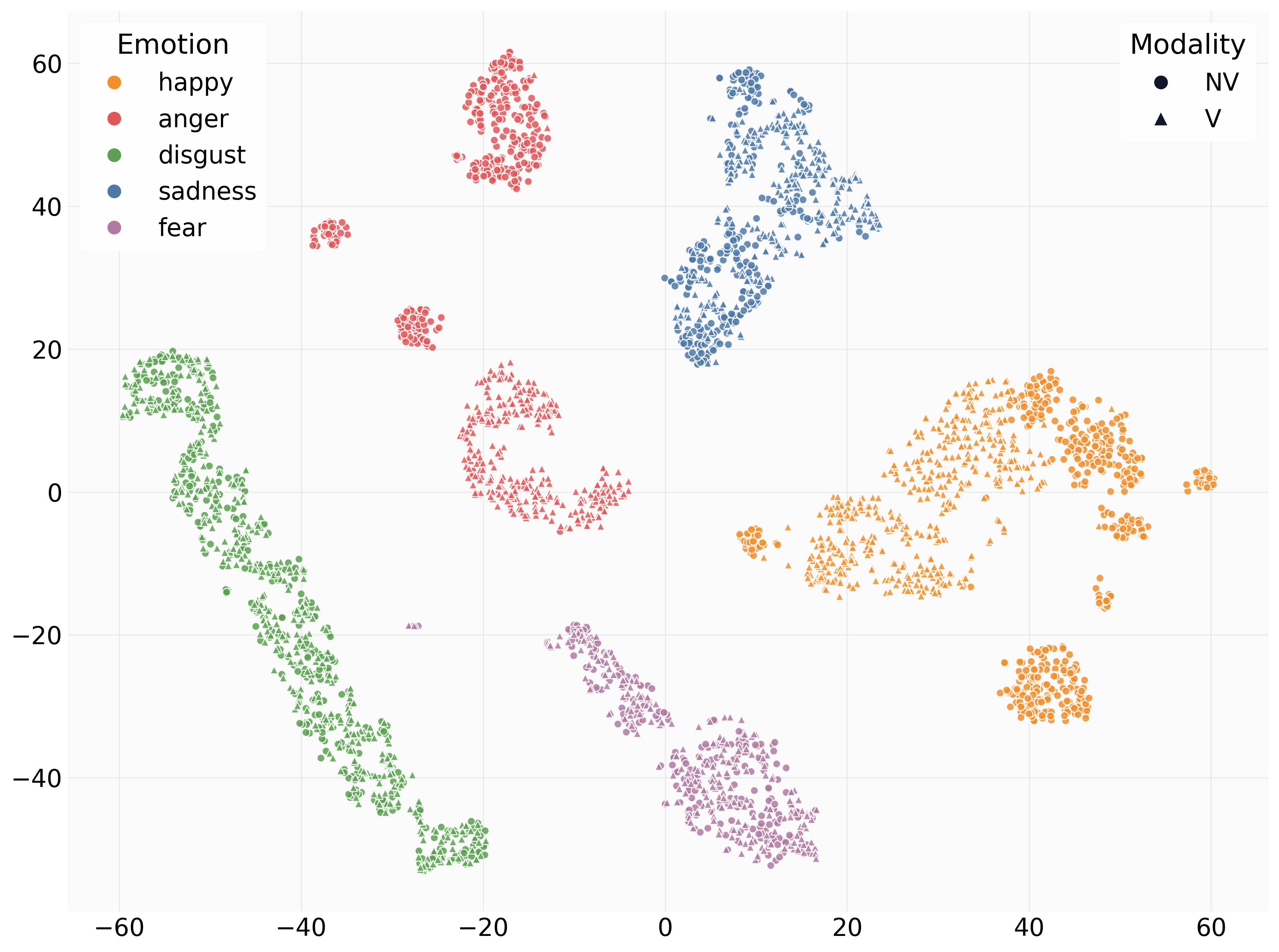}
    }\\[1ex]

    \subfloat[]{%
        \includegraphics[width=0.32\textwidth]{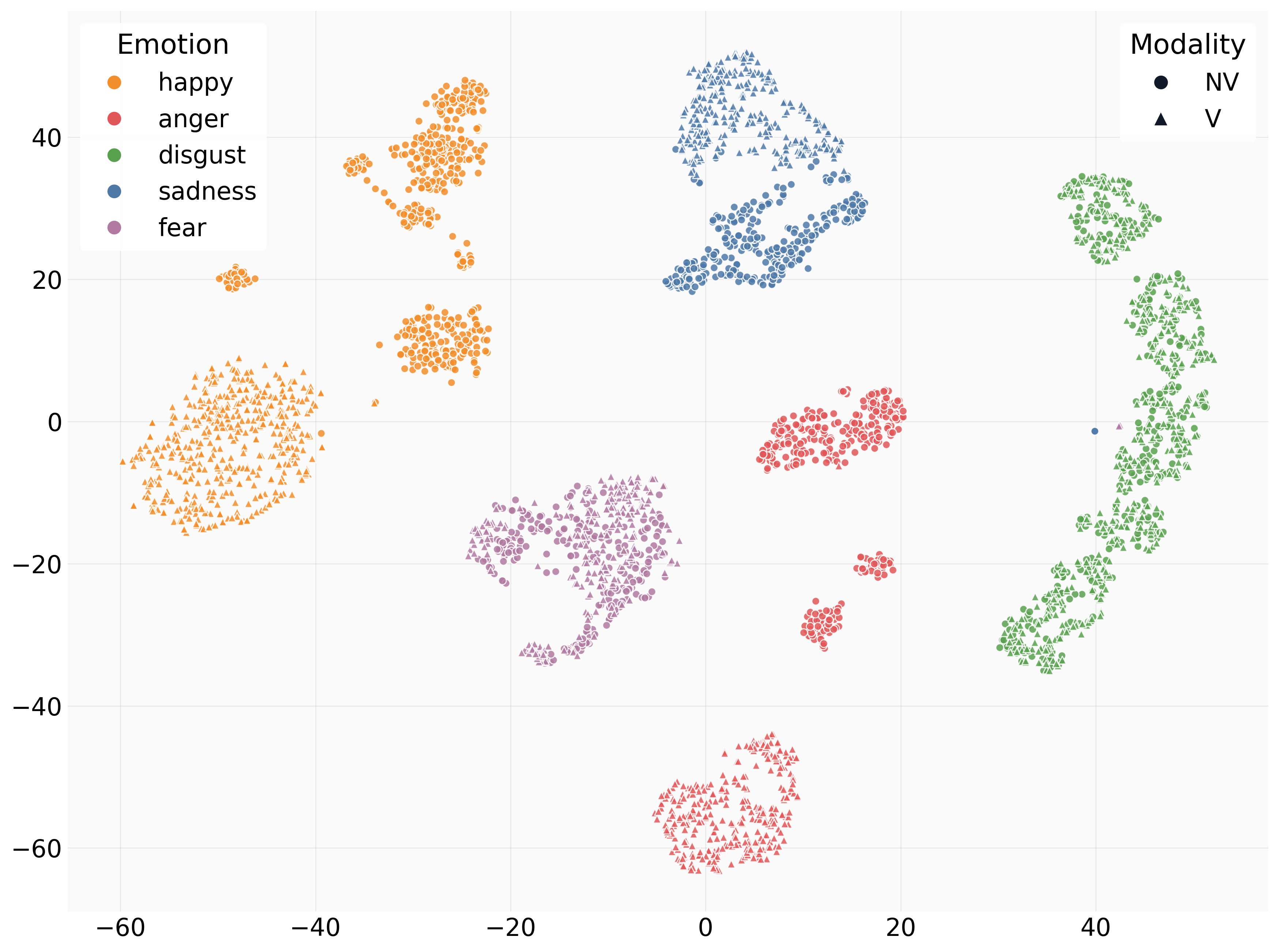}
    }
    \hspace{0.03\textwidth}
    \subfloat[]{%
        \includegraphics[width=0.32\textwidth]{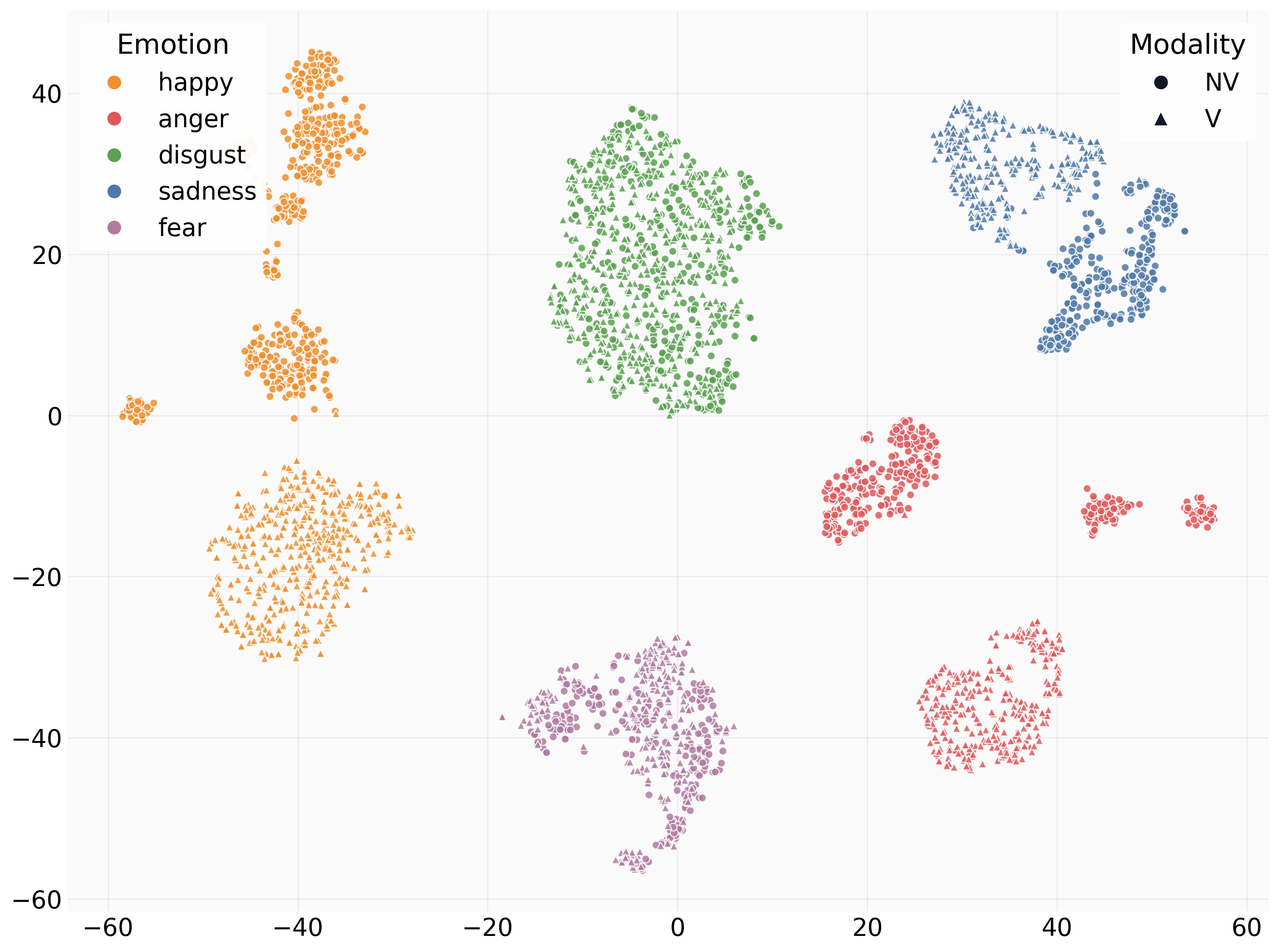}
    }
    
    \caption{Representing \textbf{NOVA-ARC} configurations. Each displays true versus predicted class distributions across the combined diagnosis and severity categories: (a) ASVP-NV WavLM; (b) ASVP-NV Voc2vec; (c) ASVP-NV Wav2vec 2.0; (d) ASVP-NV MMS; (e) \textbf{NOVA-ARC} on Voc2vec for ASVP-NV(Source)-RAVDESS(Target); (f) \textbf{NOVA-ARC} on Voc2vec for ASVP-NV(Source)-CREMA-D(Target); (g) \textbf{NOVA-ARC} on Voc2vec for ASVP-NV(Source)-MESD(Target); (h) \textbf{NOVA-ARC} on Voc2vec for ASVP-NV(Source)-AESDD(Target);. These matrices highlight classification consistency and error patterns for each fusion pairing.}
    \label{fig:tsnenv}
\end{figure*}

Figure~\ref{fig:tsnenv} visualizes the learned utterance representations using t-SNE.

\end{document}